\documentclass[11pt,a4paper]{article}

\usepackage[utf8]{inputenc}
\usepackage[T1]{fontenc}
\usepackage{jheppub}
\usepackage{amsmath,amssymb}
\usepackage{graphicx}
\usepackage[numbers,sort&compress]{natbib}
\usepackage{booktabs}
\usepackage{array}
\hypersetup{pdftitle={Phenomenology of neutrino mass ordering synergies in reactor, accelerator, and atmospheric
neutrino experiments, and the role of the detector response},pdfauthor={Vanessa Cerrone, Marco Grassi}}

\newcommand{\NO}{\mathrm{NO}}
\newcommand{\IO}{\mathrm{IO}}

\newcommand{\eVsq}{\,\mathrm{eV}^2}

\newcommand{\dmsq}[1]{\Delta m^2_{#1}}

\newcommand{\dmaa}{\Delta m^2_{\alpha\alpha}}

\newcommand{\ACP}{\mathcal{A}_{\mathrm{CP}}}

\newcommand{\avgw}[1]{\big\langle #1\big\rangle_\omega}
\newcommand{\dCP}{\delta_{\mathrm{CP}}}
\usepackage{xcolor}

\newcommand{\datm}{\delta_{\mathrm{atm}}}
\newcommand{\PNO}{P_i^{\NO}}
\newcommand{\PIO}{P_i^{\IO}}
\newcommand{\meq}[1]{\eqr{#1}}
\newcommand{\eqr}[1]{eq.~\eqref{#1}} \newcommand{\Eqr}[1]{Eq.~\eqref{#1}} \AtBeginDocument{}

\title{Phenomenology of neutrino mass ordering synergies in reactor, accelerator, and atmospheric neutrino experiments, and the role of the detector response}
\author[a]{Vanessa Cerrone}
\author[a]{and Marco Grassi}
\affiliation[a]{Dipartimento di Fisica e Astronomia dell'Universit\`a di Padova\\
and INFN Sezione di Padova, Via Marzolo 8, 35131 Padova, Italy}
\emailAdd{vanessa.cerrone@studenti.unipd.it}
\emailAdd{marco.grassi@unipd.it}

\abstract{The neutrino mass ordering is one of the main open questions of neutrino physics, addressed by
reactor, accelerator, and atmospheric experiments and by their combination. When the data are fitted under the wrong ordering, each
experiment returns its own value of $\dmsq{31}$, its wrong-ordering solution. We show that this solution is a weighted average of the values preferred by the individual bins of the measured event spectra, in reconstructed energy and, for atmospheric neutrinos, also zenith angle, with weights set by their statistical precision. Both the values and the weights depend on the detector response. In the plane of $\dmsq{31}$ fitted under the two orderings, the
solutions lie on lines of slope close to $-1$, so that each experiment is described by one
intercept. The weighted bin average extends the mass ordering sum rule to
any oscillation experiment, including atmospheric neutrinos in Earth matter. It recovers the known closed forms for short-baseline reactors and for accelerators in vacuum, and for JUNO it equals a closed form evaluated at one effective energy, which depends on the energy resolution. Above
the MSW resonance the frequency of atmospheric $\nu_\mu$ disappearance becomes nearly independent of $\dCP$, and the $\nu_\mu$ disappearance solution moves mainly with $\theta_{23}$. The atmospheric intercept depends on the energy and direction resolutions, the energy range, and the event classes, and differs between detectors. The synergy of two experiments is determined by the separation of their intercepts
relative to their precisions on $|\dmsq{31}|$. With JUNO, this separation depends strongly on
$\dCP$ for an accelerator and is nearly independent of $\dCP$ and
$\theta_{23}$ for IceCube DeepCore using all channels.}

\keywords{Neutrino Mixing, Neutrino Detectors and Telescopes (experiments), Long-Baseline Neutrino Experiments}
\arxivnumber{2611.21252}

\begin{document}
\maketitle
\flushbottom

\section{Introduction}
\label{sec:intro}
Reactor, accelerator, and atmospheric experiments determine the atmospheric mass splitting,
$\dmsq{31}$, mainly from the disappearance of $\bar\nu_e$ at reactors and of $\nu_\mu$ and $\bar\nu_\mu$ in
the other two
\cite{JUNO:2026osc,DayaBay:2022orm,NOvA:2025tmb,T2K:2023smv,T2K:2025yoy,IceCubeCollaboration:2023wtb,KM3NeT:2024flw,Super-Kamiokande:2023ahc}. Each class of experiments is sensitive to the neutrino mass ordering (NMO) in its own way.
JUNO \cite{JUNO:2015yb,JUNO:2021phys,JUNO:2024jaw} is designed to determine it from the
interference of the two fast oscillations in $\bar\nu_e$ disappearance at a medium baseline
\cite{Petcov:2001sy,Learned:2006wy,Li:2013zyd}. Accelerator experiments are sensitive through the matter effect in $\nu_e$ appearance \cite{NOvA:2025tmb,T2K:2023smv,T2K:2025wet}, and atmospheric detectors
through the Earth matter effect across the Mikheyev--Smirnov--Wolfenstein (MSW) resonance \cite{Wolfenstein:1977ue,Mikheyev:1985zog,Akhmedov:2012ah,IceCube:2023ins}.

Combining experiments adds a further source of information. Each experiment measures $\dmsq{31}$ through the oscillation frequencies of its channels. These
frequencies depend on the mass ordering in a way set by the channel, baseline, energy, and matter density of each experiment \cite{deGouvea:2005hk,Nunokawa:2005nx,Akhmedov:2012ah,Qian:2015waa}. An ordering analysis fits the data under both orderings and compares the two minima of the $\chi^2$, and the fit under the wrong ordering measures how strongly the data disfavor it. In this fit each experiment therefore returns a different value of $\dmsq{31}$, which we call its wrong-ordering solution. In a joint fit under the wrong ordering a single value of $\dmsq{31}$ has to describe all experiments, and their disagreement disfavors the wrong ordering beyond the intrinsic sensitivities of the individual experiments. This
synergy in the determination of the NMO is known for reactors combined with accelerator
\cite{Nunokawa:2005nx,Cabrera:2020ksc,Parke:2024xre,JUNO:2015yb,Li:2013zyd,Minakata:2006gq} and atmospheric experiments
\cite{Blennow:2013vta,IceCube-Gen2:2019fet,KM3NeT:2021rkn}, where it has been evaluated both in phenomenological
studies and in joint sensitivity studies of the experiments. Accelerator and atmospheric experiments have also
been analyzed jointly \cite{Ghosh:2015ena,T2K:2024wfn}. Reactor data have been combined with accelerator data \cite{JUNO:2026osc,NOvA:2025tmb,Machado:2026shm} and with atmospheric data \cite{Jin:2026evidence,Machado:2026shm}, and global fits combine all of them \cite{Esteban:2024eli,Capozzi:2025wyn,deSalas:2020pgw,Esteban:2026phq}.

We take the true ordering to be normal (NO), so that the wrong ordering is the inverted one (IO). Two experiments $A$ and $B$ that fit NO data under the IO hypothesis return their own solutions, $f_A$ and $f_B$, the fitted values of $\dmsq{31,\IO}$, with uncertainties $\sigma_A$ and $\sigma_B$. The ordering sensitivity of their combination is then~\cite{Blennow:2013vta}
\begin{equation}
\Delta\chi^2_{\rm NMO}\simeq\Delta\chi^2_A+\Delta\chi^2_B+\Delta\chi^2_{\rm syn},\qquad
\Delta\chi^2_{\rm syn}=\frac{\big(f_A-f_B\big)^2}{\sigma_A^2+\sigma_B^2}.
\label{eq:dchi2}
\end{equation}
The first two terms are the intrinsic sensitivities of the individual experiments. The third is the synergy term, which depends only on the two wrong-ordering solutions and on their precisions.

For reactor and accelerator disappearance in vacuum each channel has one fast oscillation, whose frequency is a combination of $\dmsq{31}$ and $\dmsq{21}$ that depends on the channel, and the wrong-ordering solution is known in closed form as a function of the splitting of the true ordering
\cite{deGouvea:2005hk,Nunokawa:2005nx,Minakata:2006gq,Minakata:2007tn,Parke:2016joa,Cabrera:2020ksc,Parke:2024xre}. The $\nu_\mu$ disappearance of the accelerator experiments T2K \cite{T2K:2011qtm} and NOvA \cite{NOvA:2021nfi} and the $\bar\nu_e$ disappearance of
the medium-baseline reactor JUNO were combined into a mass ordering sum rule \cite{Parke:2024xre}. The rule states that if the two channels measure the same splitting under the true ordering, their splittings under the wrong ordering differ by a quantity set by $\dmsq{21}$, the mixing angles, and $\cos\dCP$, with $\dCP$ the CP-violating phase. On the other hand, atmospheric neutrinos are
observed over many baselines and energies, across the MSW resonance,
where Earth matter effects are large. A single frequency does not describe these data, and the closed forms do not apply. For example, for the proposed atmospheric detector PINGU, numerical studies found that its wrong-ordering solution differs from the vacuum closed forms by an amount comparable to its precision on $|\dmsq{31}|$ \cite{Blennow:2013vta},
and that Earth matter enlarges this difference \cite{IceCube-Gen2:2019fet}.

In this work we describe the wrong-ordering solution of any oscillation experiment in a single way, as a weighted average of the values of $\dmsq{31,\IO}$ preferred by the individual bins of the measured event spectra, in reconstructed energy and, for atmospheric neutrinos, also zenith angle, with weights set by the statistical precision of each bin. Both the values and the weights depend on the detector response. This weighted bin average recovers the closed forms and extends the sum rule to any oscillation experiment, including atmospheric neutrinos in Earth matter. Our main application is to atmospheric neutrinos. We show how Earth matter moves the atmospheric solution away from the closed forms, how this solution depends on the unknown $\dCP$ and on the atmospheric mixing angle $\theta_{23}$, and how it changes with the detector, using IceCube DeepCore~\cite{IceCube:2011ucd} and KM3NeT/ORCA \cite{KM3NeT:2021ozk} as examples. For a medium-baseline reactor the same description gives the dependence of the solution on the energy resolution. We then compare the solutions of present reactor, accelerator, and atmospheric experiments, which set the synergy of their combinations.

Section~\ref{sec:formula} derives the weighted bin average, \autoref{sec:vacuum} recovers the reactor and accelerator closed forms and gives the dependence of JUNO on its energy resolution, \autoref{sec:atm} treats atmospheric neutrinos with DeepCore as the main example, compares it with the first six detection units of KM3NeT/ORCA (ORCA6), and tests it against a full fit, \autoref{sec:synergy} compares the solutions of current experiments with each other and with their published best fits, and \autoref{sec:comb} discusses the synergy of combinations. Section~\ref{sec:concl} summarizes the results. The inputs of each experiment, namely exposures, spectra, and detector responses, are described in appendix~\ref{sec:s_experiments}, and the derivations and numerical checks in the
other appendices. The code and the inputs that reproduce all figures and
numbers of this paper are public \cite{nmosynergies:2026}.

\section{The wrong-ordering solution as a weighted bin average}
\label{sec:formula}
Consider an experiment that measures event counts in bins $i$ of its observables, for example the
reconstructed energy for a reactor or accelerator experiment, or the reconstructed energy and zenith angle $\theta_z$ for an atmospheric detector. Each bin has an unoscillated expectation $N^{\rm u}_i$ and an oscillation
probability $P^H_i$, averaged over the true energies and directions of the events reconstructed in the bin, for the mass
ordering $H$, so that its expected count is $N^H_i=N^{\rm u}_iP^H_i$. The ordering is determined by fitting the same data under both hypotheses and comparing the two minima of the $\chi^2$. To evaluate sensitivities we use Asimov data \cite{Cowan:2010js}, the expected event counts without statistical fluctuations. On NO Asimov data the NO fit reproduces the data exactly, with $\chi^2=0$ at the true parameters, so the ordering sensitivity is the minimum of the fit under the IO hypothesis, which we call the IO fit and on which we focus.

Unless stated otherwise, the true values of the oscillation parameters are the NuFIT~6.0
best fit for normal ordering \cite{Esteban:2024eli}, listed in appendix~\ref{sec:s_notation}, and the unknown $\dCP$ and $\theta_{23}$ are varied
over their $3\sigma$ ranges where indicated. The solar parameters $\theta_{12}$ and $\dmsq{21}$ and the reactor angle
$\theta_{13}$ are fixed to these values in both orderings and in all experiments. Without superscript, $\theta_{23}$ and $\dCP$ denote the true values. The IO fit adjusts $\dmsq{31,\IO}$
and, where stated, also its own mixing angle and CP phase, $\theta_{23}^{\IO}$ and $\dCP^{\IO}$. Where they are not adjusted, $\theta_{23}^{\IO}$ equals the true $\theta_{23}$ and $\dCP^{\IO}$ is the NuFIT~6.0 IO best fit, $285^\circ$.

For NO Asimov data at the true splitting $x=\dmsq{31,\NO}$, the IO fit minimizes
\begin{equation}
\chi^2(y)=\sum_i W_i\,\big[P^{\IO}_i(y)-P^{\NO}_i(x)\big]^2,\qquad W_i=\frac{N^{\rm u}_i}{P^{\NO}_i(x)},
\label{eq:chi2}
\end{equation}
with $y=\dmsq{31,\IO}$ and $W_i$ the inverse Poisson variance of the measured probability $N_i/N_i^{\rm u}$. The IO fit returns the value of $y$ whose predicted counts are closest to the NO data, in the sense of \eqr{eq:chi2}.
Each term equals $(\Delta N_i)^2/N_i$, twice the Poisson log-likelihood ratio to second order in $\Delta N_i$
\cite{Baker:1983tu,Ji:2019rgi}. Although the fit searches for a common $y$ minimizing the total $\chi^2$, we define $y_i$ to be the value ``preferred'' by bin $i$, that is, the IO splitting that reproduces the NO data of that bin when
$P^{\IO}_i$ is expanded linearly around the solution $f$,
\begin{equation}
y_i=f+\frac{P^{\NO}_i(x)-P^{\IO}_i(f)}{s_i},\qquad s_i=\frac{\partial P^{\IO}_i}{\partial y}\bigg|_{y=f}.
\label{eq:yi}
\end{equation}
It equals the exact solution of $P^{\IO}_i(y_i)=P^{\NO}_i(x)$ when $P^{\IO}_i$ is close to linear between $f$ and $y_i$, that is, when the bin is not near an oscillation maximum or minimum, where $s_i$ vanishes. With this definition the condition $\partial\chi^2/\partial y=0$ becomes
\begin{equation}
f(x)=\frac{\sum_i\omega_i\,y_i}{\sum_i\omega_i},\qquad \omega_i=W_i\,s_i^2 .
\label{eq:master}
\end{equation}
The wrong-ordering solution is therefore the average of the values preferred by the individual bins, each weighted by the inverse variance with which it determines the splitting. We call this average the weighted bin average. Bins in which $P_i^{\IO}$ varies most steeply with $y$, that is, with the largest $|s_i|$, dominate, while bins at oscillation maxima or minima, where $s_i$
vanishes, do not contribute. Since $y_i$ and $\omega_i$ are evaluated at $f$, \eqr{eq:master} is a condition that $f$ satisfies at the minimum, exact for the $\chi^2$ of \eqr{eq:chi2}. It does not replace the fit. It splits the solution into the contributions of the bins, so that the solution for a subset of the bins, or for the same bins with other weights, follows to first order from the same $y_i$ without a new fit. It has the structure of any one-parameter least-squares minimum~\cite{ParticleDataGroup:2026mpi}, written so
that the contribution of each bin is explicit, and separates what each bin prefers, $y_i$, which is set by
the oscillation physics and by the averaging of the probability over the bin, from its weight in the average, $\omega_i$, which also depends on the detector and the event rate.

The solution depends on the true splitting $x$. In vacuum the oscillation frequency measured in each channel lies below $\dmsq{31}$ in the NO and above $|\dmsq{31}|$ in the IO, by a term proportional to $\dmsq{21}$ that does not depend on $\dmsq{31}$ \cite{deGouvea:2005hk,Nunokawa:2005nx}. The IO fit matches these frequencies, so it returns $|\dmsq{31,\IO}|=x-b$, with $b$ independent of $x$ and positive. Other effects, such as Earth matter effects, make $b$ depend slightly on $x$, but it stays positive for every experiment considered here. We therefore look for solutions of the form
\begin{equation}
f(x)=-x+b,\qquad b\equiv\dmsq{31,\IO}+\dmsq{31,\NO},
\label{eq:b_def}
\end{equation}
lines of slope close to $-1$ in the plane of the two splittings, and describe each solution by its intercept $b$. Since $\dmsq{31,\IO}$ is negative, $b=\dmsq{31,\NO}-|\dmsq{31,\IO}|$. Close to the measured value of $\dmsq{31,\NO}$ each experiment is therefore described by a single number, its intercept $b$. Two experiments share the true splitting $x$, so $f_A-f_B=b_A-b_B$, and the synergy term of \eqr{eq:dchi2} is set by the difference of their intercepts. We assign to each bin the intercept $b_i=y_i+x$, and the intercept of the experiment is their weighted average,
\begin{equation}
b=\avgw{b_i}\equiv\frac{\sum_i\omega_i\,b_i}{\sum_i\omega_i} .
\label{eq:bi}
\end{equation}
When all $y_i$ coincide the weights drop out of the average and $b$ is determined by the oscillation parameters alone, as for short-baseline reactors and accelerators in vacuum, whose solutions are known in closed form, as discussed in section~\ref{sec:vacuum}. When the bins prefer different values, as for a medium-baseline reactor and for atmospheric
neutrinos in Earth matter, $b$ depends on how the detector distributes the weights.

\Eqr{eq:master} holds for bins of any channel, and for bins that contain events of several channels. An appearance bin enters the average in the same way as a disappearance bin,
with its own $y_i$ and $\omega_i$. In current accelerator experiments the disappearance bins carry most of the weight and set the
solution, while the appearance bins mainly determine the CP phase of the IO fit. In atmospheric detectors a bin contains $\nu_e$ and $\nu_\tau$ charged-current and neutral-current events together with the $\nu_\mu$ charged-current events. These events add to the expected count of the bin, which changes its $y_i$ and $\omega_i$ and can move the solution.

When the IO fit adjusts only $\dmsq{31,\IO}$, we obtain the solution by solving \eqr{eq:master} iteratively, because $y_i$ and $\omega_i$ depend on $f$. This iteration
is the Gauss--Newton minimization of the $\chi^2$ of \eqr{eq:chi2}, so that the weighted bin average and the minimum of
the $\chi^2$ coincide. When the IO fit also adjusts other parameters, such as its mixing angle and CP phase, \eqr{eq:master} holds at their fitted values. Appendix~\ref{sec:s_formula} gives the derivation of \eqr{eq:master}, its form with the orderings exchanged and with nuisance parameters, and the minimization.

The detector response, that is, how the detector maps the true energies and directions of the neutrinos onto the analysis bins, enters \eqr{eq:master} in two ways. First, the probability $P_i$ is averaged over the true energies and
directions of the events reconstructed in the bin, so the energy and direction resolutions determine which oscillation phases are averaged together. Second, the weights $\omega_i$ depend on the event rate and on how sharply each bin resolves the
oscillation pattern.

\begin{figure}[t]\centering
\includegraphics[width=\textwidth]{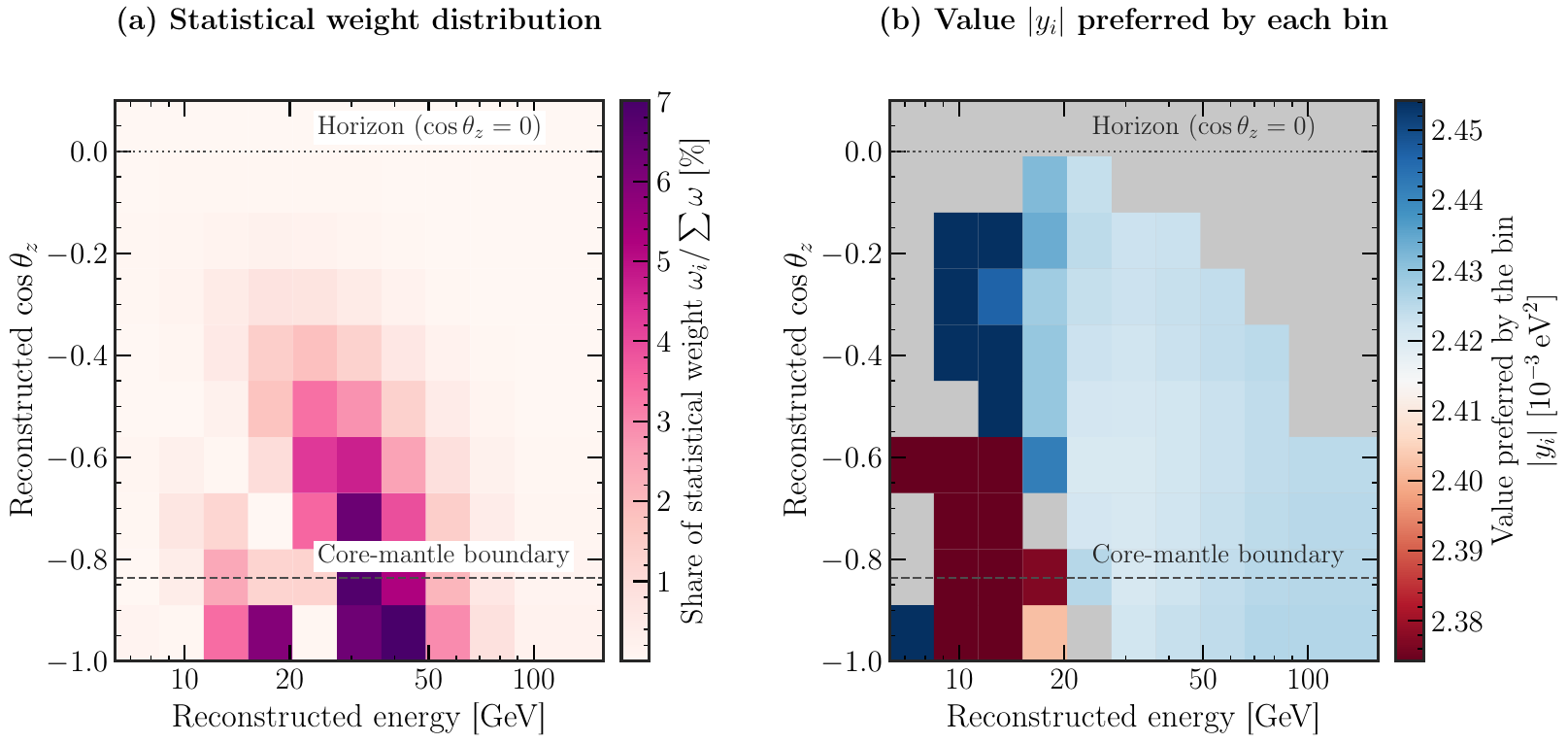}
\caption{The weighted bin average for IceCube DeepCore at the NuFIT~6.0 best
fit, on the response of the public data release ($\nu_\mu$ disappearance, with $\theta_{23}$ and the CP phases
fixed at their NuFIT~6.0 values). (a) Share of
the total statistical weight $\omega_i$ carried by each reconstructed bin. (b) The value $|y_i|$ preferred by each reconstructed bin, \eqr{eq:yi}, where a smaller $|y_i|$ is a larger bin intercept $b_i=x-|y_i|$ (bins with negligible weight in gray, color scale centered on the solution and saturated). Dashed and dotted reference lines indicate the core--mantle boundary
($\cos\theta_z=-0.838$) and the horizon ($\cos\theta_z=0$). }
\label{fig:weights}
\end{figure}

Figure~\ref{fig:weights} shows an example, the weights and the values $y_i$ of DeepCore,
computed with the detector response of its public data release, described in \autoref{sec:atm}. About two thirds of the weight lies between 20 and 50~GeV for up-going neutrinos, in bins that prefer nearly the same value. Below 20~GeV, the bins whose neutrinos cross the core or the deep mantle carry about a sixth of the weight and prefer a much smaller $|\dmsq{31,\IO}|$, that is, a larger bin intercept $b_i$. The intercept of the experiment therefore lies above the intercepts of the bins that carry most of the weight.

The $\chi^2$ of \eqr{eq:chi2} is a simplified version of the one an experiment minimizes. It fits only the oscillation parameters and contains no systematic uncertainties. Evaluating it on Asimov data is the standard choice for sensitivity studies, and the $\Delta\chi^2$ obtained in this way approximates the median outcome of real data \cite{Blennow:2013oma,Qian:2012zn,Ciuffoli:2013rza}. Appendix~\ref{sec:s_valid} compares this simplified $\chi^2$ with a full fit of the DeepCore public data. The weighted bin average keeps its form when systematic uncertainties are included through nuisance parameters that are profiled in the fit~\cite{Fogli:2002pt}. As derived in appendix~\ref{sec:s_nuis}, to first order the values $y_i$ preferred by the bins do not change, and the nuisance parameters only redistribute the weight among the bins. They therefore move the solution only through the spread of the $y_i$, which vanishes for short-baseline reactors and accelerators in vacuum and is set by the solar phase for JUNO and by Earth matter for atmospheric detectors. If $\theta_{13}$, fixed in this work, were fitted with an external prior, such as that of Daya Bay, it would act in the same way.

\section{Reactor and accelerator experiments: the closed forms as limits}
\label{sec:vacuum}
\subsection{Values preferred by the bins in vacuum}
By \eqr{eq:bi} the intercept of an experiment is the weighted average of the bin intercepts $b_i$. For reactor and accelerator experiments matter effects are small, so we first derive $b_i$ in vacuum. Where all $y_i$ coincide, the weights drop out of the average and a closed form follows. In vacuum the survival probability of flavor
$\alpha=e,\mu$ can be written with one fast oscillation \cite{Nunokawa:2005nx,Minakata:2006gq,Minakata:2007tn,Parke:2016joa},
\begin{equation}
P_{\alpha\alpha}=1-4u_1u_2\sin^2\Delta_{21}-2u_3(1-u_3)\big[1-C_{\alpha\alpha}\cos(2|\Delta_{\alpha\alpha}|\pm\varphi_{\alpha\alpha})\big],
\label{eq:twofreq_main}
\end{equation}
where $u_k=|U_{\alpha k}|^2$ are the squared moduli of the elements of row $\alpha$ of the lepton mixing matrix $U$, $\Delta_{ij}=\dmsq{ij}L/4E$, and the
sign is $+$ for the normal and $-$ for the inverted ordering. Three quantities describe the fast oscillation. Its frequency is the effective splitting
\begin{equation}
\dmsq{\alpha\alpha}=\dmsq{31}-\kappa_\alpha\,\dmsq{21},\qquad
\kappa_\alpha=\frac{|U_{\alpha2}|^2}{|U_{\alpha1}|^2+|U_{\alpha2}|^2},
\label{eq:dmaa}
\end{equation}
with phase $\Delta_{\alpha\alpha}=\dmsq{\alpha\alpha}L/4E$. Its amplitude $C_{\alpha\alpha}$, given in appendix~\ref{sec:s_vacuum}, is close to one when the solar oscillation has not yet developed. Its phase is shifted by the solar phase
\begin{equation}
\varphi_{\alpha\alpha}=\arctan\!\big[(1-2\kappa_\alpha)\tan\Delta_{21}\big]-(1-2\kappa_\alpha)\,\Delta_{21},
\label{eq:phi_main}
\end{equation}
taken on the branch of the arctangent that is continuous in $\Delta_{21}$~\cite{Parke:2016joa}. It enters with opposite signs in the two orderings. It vanishes for small $L/E$, starts at the third power of $\Delta_{21}$, and its size grows with $L/E$~\cite{Minakata:2007tn,Parke:2008cz}. The derivation of the formulas of this section is in appendix~\ref{sec:s_vacuum}.

Consider first a bin described by a single energy $E_i$. The IO probability of the bin equals the NO one when the fast oscillations of \eqr{eq:twofreq_main} are in phase in the two orderings~\cite{deGouvea:2005hk,Nunokawa:2005nx,Minakata:2007tn,Qian:2012xh,Qian:2015waa}. Since the solar phase enters with opposite signs, this condition reads
\begin{equation}
|\Delta_{\alpha\alpha}|^{\IO}=|\Delta_{\alpha\alpha}|^{\NO}+\varphi_{\alpha\alpha}(E_i).
\label{eq:phase_match}
\end{equation}
The two phases $\Delta_{\alpha\alpha}$ are proportional to $L/E$. The solar phase is not, since the part of the solar oscillation linear in $L/E$ is contained in the frequency through $\kappa_\alpha$~\cite{Parke:2016joa}. Multiplying \eqr{eq:phase_match} by $4E_i/L$ and inserting \eqr{eq:dmaa} for the two orderings gives the value preferred by the bin in vacuum,
\begin{equation}
y_i=-x+\big(\kappa_\alpha^{\NO}+\kappa_\alpha^{\IO}\big)\dmsq{21}-\Delta m^2_{\varphi,\alpha\alpha}(E_i),\qquad
\Delta m^2_{\varphi,\alpha\alpha}\equiv\frac{4E\,\varphi_{\alpha\alpha}}{L},
\label{eq:yb_vac}
\end{equation}
where $\kappa_\alpha$ is evaluated with $\dCP$ and $\theta_{23}$ of each ordering. The bin intercept is a term common to all bins, set by the mixing, minus the solar-phase term of the bin, which depends on energy because $\varphi_{\alpha\alpha}$ is not linear in $L/E$~\cite{Parke:2016joa}.

When the solar phase is negligible, a single value of $y$ puts the two fast oscillations in phase at every energy at once. Every bin then prefers this value, whatever its energy, its width, or the smearing by the energy resolution, since these only average the probability over energies that all prefer the same value. The detector enters \eqr{eq:master} only through the weights, which cannot move an average of equal values, and the solution is a closed form set by the oscillation parameters alone.

This no longer holds when the two orderings differ by a phase that is not proportional to $L/E$. For a medium-baseline reactor the solar-phase term depends on energy, as shown for JUNO in \autoref{sec:reactors}. For atmospheric neutrinos Earth matter shifts the effective splittings of the two orderings by amounts that depend on the energy and on the path through the Earth, as discussed in \autoref{sec:atm}. The bins then prefer different values, and the weights, which carry the detector response, set where their average lies.

\subsection{Accelerators}
For accelerator disappearance ($\alpha=\mu$) at long baselines (LBL), $\varphi_{\mu\mu}$ is negligible \cite{Nunokawa:2005nx}, so all $y_i$ coincide. As a consequence the solution does not depend on the binning or on the detector, and
\eqr{eq:master} reduces to
\begin{equation}
f_{\rm LBL}(x)=-x+(\kappa_{\mu}^{\NO}+\kappa_{\mu}^{\IO})\,\dmsq{21},\qquad
\kappa_\mu\simeq c_{12}^2-\ACP,
\label{eq:lbl}
\end{equation}
where $\ACP=\cos\dCP\sin\theta_{13}\sin2\theta_{12}\tan\theta_{23}$ carries the leading dependence on the CP phase \cite{Nunokawa:2005nx,Parke:2024xre}, with $s_{ij}=\sin\theta_{ij}$ and $c_{ij}=\cos\theta_{ij}$. We call $-(\ACP^{\NO}+\ACP^{\IO})\dmsq{21}$, the leading part of the intercept that depends on the CP phases, the CP term. From \eqr{eq:dmaa}, the exact coefficient is
\begin{equation}
\kappa_\mu=\frac{c_{12}^2c_{23}^2+s_{12}^2s_{23}^2s_{13}^2-2s_{12}c_{12}s_{23}c_{23}s_{13}\cos\dCP}{1-c_{13}^2s_{23}^2} ,
\label{eq:kmu_exact}
\end{equation}
and the second relation in \eqr{eq:lbl} is its expansion to first order in $\sin\theta_{13}$, which makes the dependence on $\dCP$ explicit. All numerical results use the exact form. With the NuFIT~6.0 values, and $\sin^2\theta_{23}=0.561$ in both orderings, the intercept of \eqr{eq:lbl} is
\begin{equation}
b_{\rm LBL}\simeq\big[103.8-11.6\,(\cos\dCP+\cos\dCP^{\IO})\big]\times10^{-6}\eVsq ,
\label{eq:lbl_num}
\end{equation}
where $\dCP$ is the true phase and $\dCP^{\IO}$ that of the IO fit.

\begin{figure}[t]\centering
\includegraphics[width=0.49\textwidth]{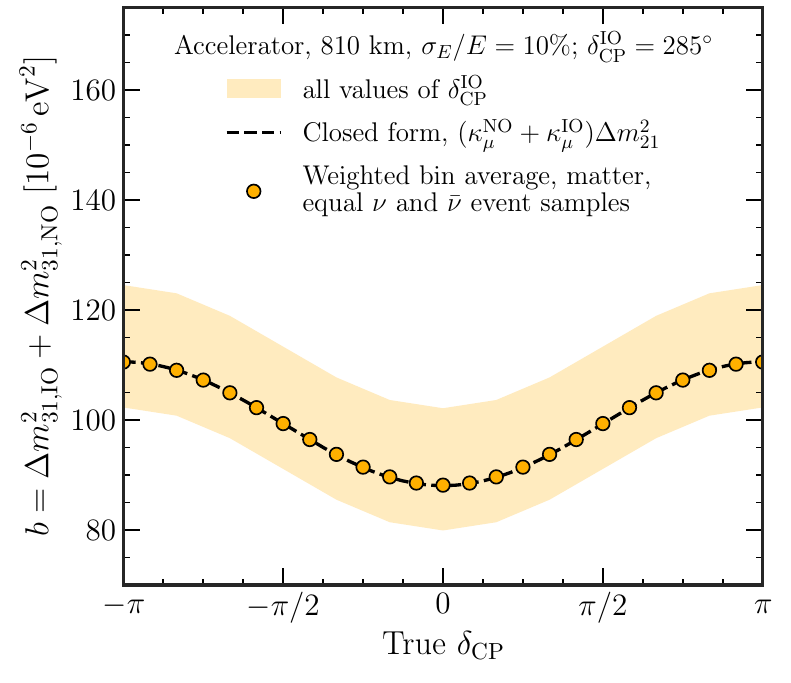}\hfill
\includegraphics[width=0.49\textwidth]{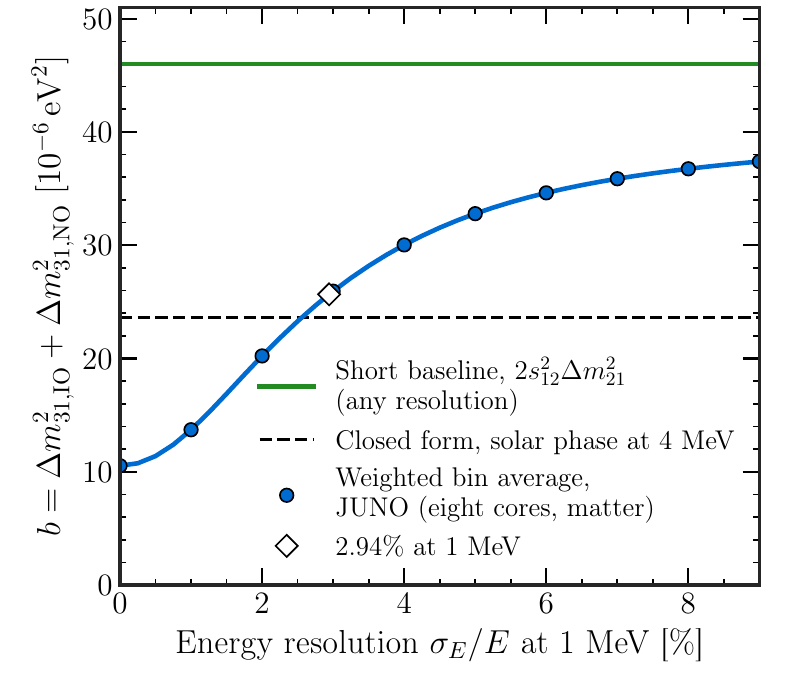}
\caption{Left: accelerator disappearance (810~km, $\sigma_E/E=10\%$, Earth's
crust, equal $\nu$ and $\bar\nu$ event samples). The intercept from
\eqr{eq:master} (points) against the closed form \eqref{eq:lbl} (dashed line),
as a function of the true $\dCP$ with $\dCP^{\IO}$ at its NuFIT~6.0 value. The points and the line agree to $0.1\times10^{-6}\eVsq$. The shaded range covers all values of $\dCP^{\IO}$, which give nearly equally good solutions in
disappearance. Right: JUNO, eight reactor cores, in matter. The
intercept from \eqr{eq:master} as a function of the energy resolution $\sigma_E/E$ at 1~MeV, with the energy dependence of the resolution of the JUNO simulation \cite{JUNO:2024reso} scaled by a common factor, and at the resolution of this simulation, $2.94\%$ at 1~MeV (open diamond). The short-baseline intercept and \eqr{eq:mbr} with the solar-phase term evaluated at 4~MeV in place of its weighted average (dashed) are shown for comparison. The dashed line agrees with the weighted bin average only near the predicted resolution of JUNO.
Oscillation parameters from NuFIT~6.0, with $\dmsq{31,\NO}=2.534\times10^{-3}\eVsq$.}
\label{fig:vacuum}
\end{figure}

The left panel of \autoref{fig:vacuum} compares the weighted bin average with this closed form for an 810~km beam with equal neutrino and antineutrino event samples, against the true $\dCP$. The two agree over the full range of $\dCP$. The band shows the range of the
intercept over all values of $\dCP^{\IO}$, which disappearance alone does not determine. A change of $\dCP^{\IO}$ changes $\kappa_\mu^{\IO}$, the IO fit compensates it with a change of $\dmsq{31,\IO}$, and every value of $\dCP^{\IO}$ gives nearly the same $\chi^2$. The $\chi^2$ is therefore nearly constant along a curve in the plane of $\dCP^{\IO}$ and $\dmsq{31,\IO}$, which we call the \emph{flat valley}. A fit of the NOvA $\nu_\mu$ disappearance sample alone indeed gives a flat posterior of $\dCP$ in both orderings \cite{NOvA:Nu2026}.
The intercepts of the left panel include the matter effect of the Earth's crust, which shifts the splitting measured in $\nu_\mu$ disappearance slightly
\cite{Denton:2024thm}. The matter potential changes sign between neutrinos and antineutrinos, and we find that it
shifts the accelerator solution in opposite directions for the two. With equal neutrino and antineutrino event samples the two shifts cancel in the weighted average. With neutrinos alone they do not cancel, and the solution moves from the closed form by a few
$10^{-6}\eVsq$ at the baseline of NOvA and by less at that of T2K. Beams that also run with antineutrinos give smaller shifts. All these shifts are small compared with the range spanned by $\dCP$.

\subsection{Reactors}
\label{sec:reactors}
For reactor disappearance ($\alpha=e$), $\kappa_e=s_{12}^2$ in both orderings and does not depend on
$\dCP$. At short baselines, such as those of Daya Bay \cite{DayaBay:2022orm}, the solar oscillation
has not yet developed, the solar phase is negligible, and all $y_i$ coincide. The solution of a short-baseline reactor (SBR),
\begin{equation}
f_{\rm SBR}(x)=-x+2s_{12}^2\dmsq{21}
\label{eq:sbr}
\end{equation}
then follows from \eqr{eq:master} for any choice of weights.

At a medium baseline the solar oscillation has developed and $\Delta m^2_{\varphi,ee}$ depends
on energy \cite{Petcov:2001sy,Learned:2006wy,Zhan:2008id}. The bins prefer different values, as found for the energy ranges of JUNO in ref.~\cite{Choubey:2022gzv}, and
\eqr{eq:master} gives the solution of a medium-baseline reactor (MBR),
\begin{equation}
f_{\rm MBR}(x)\mathrel{\simeq}-x+2s_{12}^2\dmsq{21}-\avgw{\Delta m^2_{\varphi,ee}} .
\label{eq:mbr}
\end{equation}
The medium-baseline solution is the short-baseline one minus the solar-phase term averaged with the weights $\omega_i$ of the bins, so that it depends on the energy resolution and on the energy spectrum of the detected events.

The right panel of \autoref{fig:vacuum} shows the intercept of JUNO \cite{JUNO:2024jaw,JUNO:2022osc,JUNO:2024reso} from \eqr{eq:master} as a function of the energy resolution $\sigma_E/E$ at 1~MeV, together with the short-baseline intercept and a closed form with the solar phase
at a single energy, 4~MeV as in refs.~\cite{Qian:2012xh,Cabrera:2020ksc}. Our model of JUNO, described in appendix~\ref{sec:s_reactors}, includes its eight reactor cores and, although small, the matter effects of the crust \cite{Li:2016txk,Khan:2019doq}. The solar-phase term grows as the solar oscillation develops \cite{Minakata:2007tn,Parke:2008cz}, and its exact form is given in appendix~\ref{sec:s_vacuum}. At the baseline of JUNO it is largest near 2.5~MeV, close to the maximum of the solar oscillation at about 3~MeV, and it falls steeply at higher energies, where the solar oscillation is less developed. At low energy the fast oscillation also has the shortest period in energy. A worse energy resolution averages the fast oscillation in these bins first, which reduces their derivative $s_i$ and therefore their weight $\omega_i$. The bins at higher energies, where the solar-phase term is small, then carry a larger share of the weight, and the solution moves toward the short-baseline one. At the energy resolution predicted by the full simulation of JUNO \cite{JUNO:2024reso} and used in its ordering sensitivity study \cite{JUNO:2024jaw}, $2.94\%$ at 1~MeV, the weighted bin average gives $b=25.7\times10^{-6}\eVsq$. Between $2\%$ and $4\%$ at 1~MeV the intercept changes by about
$10\times10^{-6}\eVsq$. The ordering sensitivity of JUNO is known to depend on the energy resolution
\cite{JUNO:2024jaw,Qian:2012xh,Li:2013zyd}, and so does its wrong-ordering solution, by small amounts compared with its separation from the accelerator and atmospheric solutions. A combined study of JUNO and ORCA also found that the JUNO best fits in the two orderings change little with the resolution \cite{KM3NeT:2021rkn}.

Above its maximum the solar-phase term decreases with energy, so its weighted average in \eqr{eq:mbr} equals its value at one effective energy, which depends on the resolution. It is about 4~MeV at the predicted resolution of JUNO, so the closed form at 4~MeV reproduces \eqr{eq:mbr} only near this resolution. The width of the energy bins matters little as long as the bins are narrower than the resolution, as for the 20~keV bins used here.

In a full fit, systematic uncertainties such as the nonlinearity of the energy scale and the shape of the reactor
flux are mildly correlated with $\dmsq{31}$ \cite{JUNO:2022osc,JUNO:2024jaw}. They redistribute the weight among the bins, and their effect on the intercept can be evaluated with the full systematic model of the experiment \cite{Qian:2012xh,Forero:2021lax}, but it is not included in this work.

In summary, the solutions of a short-baseline reactor and of an accelerator in vacuum are determined by the oscillation parameters alone, while the solution of a medium-baseline reactor also depends on its energy resolution. The difference between the intercepts of the accelerator and of the reactor is the difference between the wrong-ordering splittings of the two channels in the mass ordering sum rule \cite{Parke:2024xre}.

\section{Atmospheric neutrinos: matter effects and detector response}
\label{sec:atm}
Atmospheric neutrinos reach the detector from all directions, along baselines from a few tens of kilometers
for down-going neutrinos to the diameter of the Earth for up-going ones, with energies from a few GeV to hundreds of GeV.
Along the paths that cross the Earth, neutrinos in the NO and antineutrinos in the IO undergo the MSW resonance of $\theta_{13}$, where the mixing angle $\theta_{13}$ in matter becomes maximal, at about 6--10~GeV in the mantle. These matter effects differ between the two orderings and give
atmospheric detectors their own sensitivity to the NMO
\cite{Akhmedov:2006hb,Akhmedov:2012ah,Winter:2013ema,Capozzi:2015bxa,Super-Kamiokande:2023ahc,Banuls:2001zn,Bernabeu:2003yp,Petcov:2005rv,Gandhi:2007td,Mena:2008rh}.

All atmospheric probabilities are computed with three flavors through the Preliminary Reference Earth Model (PREM) \cite{Dziewonski:1981xy}, with the algorithm of NuFast-Earth \cite{Denton:2025nufastearth}. Unless stated otherwise, sections~\ref{sec:atm_shift} to~\ref{sec:atm_det} use $\nu_\mu$ disappearance with an IO fit that adjusts only $\dmsq{31,\IO}$, which isolates the effect of matter on the frequency. In this reference configuration the intercept of DeepCore is $b=119.7\times10^{-6}\eVsq$. The descriptions of DeepCore and ORCA6 are in appendix~\ref{sec:s_experiments}, the derivation of the formulas of this section in appendix~\ref{sec:s_atm}, and the comparison with the full fit in appendix~\ref{sec:s_valid}.

\subsection{Matter shift of the values preferred by the bins}
\label{sec:atm_shift}
Without matter effects, each bin of an atmospheric $\nu_\mu$ sample behaves like an accelerator measurement at its own
baseline and energy, as described by \eqr{eq:yb_vac}. The solar phase $\varphi_{\mu\mu}$ is negligible at the
energies where the weights of DeepCore are largest, above about 10~GeV, so that all $y_i$ coincide and \eqr{eq:master}
returns the accelerator closed form of \eqr{eq:lbl}. When matter effects are included, the $y_i$ values differ between the bins. We define the difference $\delta_i$ between the bin intercept and the vacuum value through $b_i=2c_{12}^2\dmsq{21}+\delta_i$, where $2c_{12}^2\dmsq{21}$ is the vacuum value without the CP term, to first order in $\sin\theta_{13}$. The intercept of the solution for $\dmsq{31,\IO}$ is then this vacuum value plus the weighted average $\datm$ of the differences,
\begin{equation}
f_{\rm atm}(x)=-x+2c_{12}^2\dmsq{21}+\datm(x),\qquad \datm=\avgw{\delta_i}.
\label{eq:atm}
\end{equation}
The difference $\delta_i$ contains the matter effects, the CP term, the solar phase, and, when they are included, the contribution of the channels other than $\nu_\mu$ disappearance. In \autoref{fig:weights}(b) the $y_i$ values preferred by the DeepCore bins spread over a wide range. This is the spread of the $\delta_i$, and their weighted average $\datm$ raises the intercept above $2c_{12}^2\dmsq{21}$.

When the energy of a bin is above the resonance, $\delta_i$ has a closed form. In matter the $\nu_e$ component of the mass eigenstates
is suppressed above the resonance, but a small admixture remains. This admixture shifts the effective
oscillation frequency of $\nu_\mu$ at second order in $\sin\theta_{13}$
\cite{Akhmedov:2004ny,Minakata:2015gra,Denton:2016wmg}, in opposite directions for neutrinos and
antineutrinos and for the two orderings. Matching the shifted frequencies of the two orderings gives the difference $\delta_i$ of each bin, which we derive in appendix~\ref{sec:s_shift},
\begin{equation}
\delta_i\simeq\frac{2s_{13}^2\cos2\theta_{12}}{c_{13}^2}\dmsq{21}\pm\frac{2s_{13}^2}{c_{13}^2}\,x\,\frac{\hat A_i}{\hat A_i^2-1},
\qquad\hat A_i=\frac{2\sqrt2G_FN_{e,i}E_i}{x},
\label{eq:shift_main}
\end{equation}
with $+$ for neutrinos and $-$ for antineutrinos, to leading order in $s_{13}^2$. Here $G_F$ is the Fermi constant, and $E_i$ and $N_{e,i}$ are a true neutrino energy and the electron density averaged along its path. A reconstructed bin collects many true energies and directions, and its $\delta_i$ combines their values of \eqr{eq:shift_main}, each weighted by how strongly it changes the bin count with the splitting. The second term, the matter term, is proportional to $s_{13}^2$ and grows toward the resonance, where $\hat A_i$ is close to one and the expansion no longer holds.

A detector that does not
distinguish neutrinos from antineutrinos averages the two shifts, weighted by the shares of neutrinos and
antineutrinos in each bin. The average does not vanish, because the atmospheric flux and the cross section are larger for neutrinos than for antineutrinos. The same neutrino excess gives such a detector its own ordering sensitivity \cite{Akhmedov:2012ah,Capozzi:2015bxa}. Averaged with the weights of DeepCore and evaluated with the exact eigenvalues of appendix~\ref{sec:s_shift} in place of the expansion of \eqr{eq:shift_main}, the frequency shift accounts for about $70\%$ of $\datm$, and most of the rest comes from the amplitude of the oscillation near the resonance. \Eqr{eq:shift_main} shows how $\datm$ scales with $\theta_{13}$ and
the energy, and gives the sign of its dependence on the density.
The matter term of \eqr{eq:shift_main} depends on $x$ through its prefactor and through $\hat A_i$, so $\datm$ depends slightly on $x$, and the slope of the solution differs from $-1$ by about $1\%$ for DeepCore.

\subsection{Dependence on \texorpdfstring{$\dCP$ and $\theta_{23}$}{delta CP and theta23}}
\label{sec:atm_cp}
\begin{figure}[t]\centering
\includegraphics[width=0.66\textwidth]{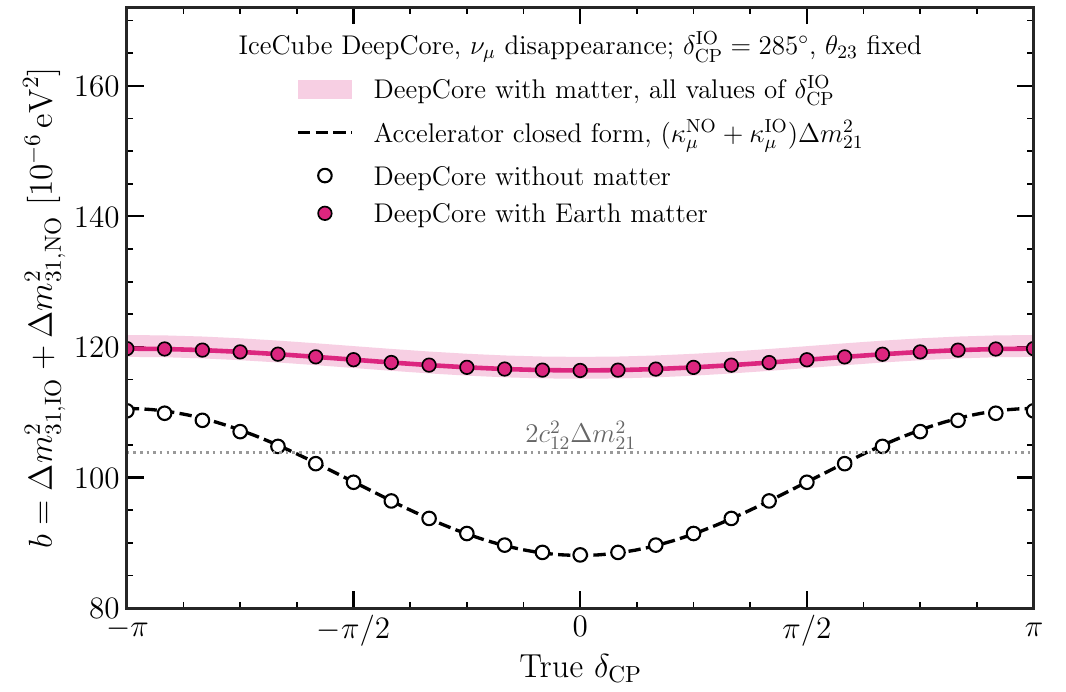}
\caption{Intercept of the IceCube DeepCore wrong-ordering solution against the true $\dCP$, computed with
Earth matter (filled points) and without (open points), together with the accelerator closed form
$(\kappa_\mu^{\NO}+\kappa_\mu^{\IO})\dmsq{21}$ (dashed line). Points and line use $\dCP^{\IO}$ at its NuFIT~6.0 value, and the band covers all values of $\dCP^{\IO}$, with matter. In both orderings $\theta_{23}$ is fixed at its true value, and the other parameters are at NuFIT~6.0, with the detector response of the IceCube DeepCore public data release \cite{IceCube:B4RITM} for $\nu_\mu$ disappearance. Dotted: $2c_{12}^2\dmsq{21}$, the vacuum value without the CP term. }
\label{fig:cp}
\end{figure}

Two sets of parameters enter the atmospheric wrong-ordering solution. The true $\dCP$ and $\theta_{23}$ define the NO data and are unknown, while $\dCP^{\IO}$ and $\theta_{23}^{\IO}$ belong to the IO fit, which adjusts them or keeps them fixed, as stated in each case. Besides shifting the values preferred by the bins, the matter effect changes how the solution depends on $\dCP$. Figure~\ref{fig:cp} shows the intercept of
DeepCore against the true $\dCP$, computed with and without Earth matter, and compares it with the accelerator
closed form. Without matter the intercepts (open points) lie on the closed form and follow $\cos\dCP$, as for accelerator
experiments. With matter (filled points) they lie higher, as found in ref.~\cite{IceCube-Gen2:2019fet}, and are nearly flat. For the same reason the band over all values of $\dCP^{\IO}$ is narrow.

The explanation lies in the decoupling of $\nu_e$ above the MSW resonance. In vacuum, $\nu_\mu$ oscillates with
the effective frequency $\dmsq{\mu\mu}$ of \eqr{eq:dmaa}, set by $\kappa_\mu$, which depends on $\dCP$. A change of $\dCP$
therefore changes the value preferred by every bin by the same amount. Above the MSW resonance $\nu_e$ decouples
\cite{Akhmedov:2004ny,Akhmedov:2012ah,Minakata:2015gra}, and $\nu_\mu$ oscillates with a single frequency of the
$\nu_\mu$--$\nu_\tau$ system. This frequency is a difference of eigenvalues of the Hamiltonian in matter, which do not
depend on $\dCP$, because $\dCP$ can be removed by a rephasing that commutes with the matter potential
\cite{Kimura:2002wd,Yokomakura:2002av,Akhmedov:2008qt,Denton:2016wmg}. Above the resonance, the value preferred by each bin is then $y_i\simeq-x+2c_{12}^2\dmsq{21}+\delta_i$, where $\delta_i$ takes the form of \eqr{eq:shift_main}, which does not contain $\dCP$. To first order in $s_{13}\dmsq{21}/|\dmsq{31}|$, the phase $\dCP$ enters only the amplitude of the oscillation, through a term proportional to $\cos2\theta_{23}$ that vanishes at
maximal mixing, $\theta_{23}=45^\circ$ \cite{Akhmedov:2008qt,Akhmedov:2012ah}. This holds for samples whose statistical weight lies mostly above the MSW resonance, as for DeepCore.
The atmospheric solution therefore depends little on $\dCP$, while the accelerator solution follows $\cos\dCP$. As a result, the distance between the atmospheric and the accelerator intercepts, which sets the synergy of an atmospheric and an accelerator experiment, changes with the true $\dCP$, as the distance between the filled points and the line in \autoref{fig:cp} shows.

For $\theta_{23}$ the situation is reversed: the accelerator solution depends weakly on $\theta_{23}$, and the atmospheric $\nu_\mu$ disappearance solution strongly. In $\nu_\mu$ disappearance $\theta_{23}$ sets the amplitude of the oscillation, $4|U_{\mu3}|^2(1-|U_{\mu3}|^2)$ with $|U_{\mu3}|^2=c_{13}^2\sin^2\theta_{23}$. It enters the frequency only through the small CP term of \eqr{eq:lbl}. In vacuum a change of $\theta_{23}^{\IO}$ cannot compensate a change of frequency, so the IO fit keeps the amplitude of the true oscillation. The true $\theta_{23}$ and its octant mirror, $|U_{\mu3}|^2\to1-|U_{\mu3}|^2$, give the same amplitude, and the IO fit returns one of the two.

In Earth matter, near the MSW resonance, part of the $\nu_\mu$ convert to $\nu_e$. This conversion grows with $\sin^2\theta_{23}$, and it occurs in neutrinos for the normal ordering and in antineutrinos for the inverted one. It gives atmospheric detectors their sensitivity to the octant of
$\theta_{23}$ \cite{Choubey:2013xqa,Choubey:2005zy}. The IO fit reduces the mismatch of this conversion between the orderings with a value of $\theta_{23}^{\IO}$ different from the true $\theta_{23}$. The pull on $\theta_{23}^{\IO}$ comes mainly from the low-energy bins whose neutrinos cross the core or the deep mantle, and it moves $\theta_{23}^{\IO}$ toward maximal mixing, which deepens the oscillation. The bins that carry most of the weight lie above the energy of the first oscillation minimum, where both a deeper oscillation and a larger $|\dmsq{31,\IO}|$ reduce the expected counts, so the IO fit compensates the deeper oscillation with a smaller $|\dmsq{31,\IO}|$. When the IO fit adjusts $\theta_{23}^{\IO}$ and $\dCP^{\IO}$, the intercept of DeepCore $\nu_\mu$ disappearance therefore rises, by an amount that depends on the true $\theta_{23}$, so the solution moves with it. Section~\ref{sec:synergy} shows this for DeepCore: against the true $\sin^2\theta_{23}$ the intercept changes little up to maximal mixing and grows in the upper octant. When all channels are included, the $\nu_e$ and $\nu_\tau$ charged-current events constrain $\theta_{23}^{\IO}$, and the solution depends much less on the true $\theta_{23}$.

\subsection{Detector response}
\label{sec:atm_det}
\begin{figure}[t]\centering
\includegraphics[width=0.49\textwidth]{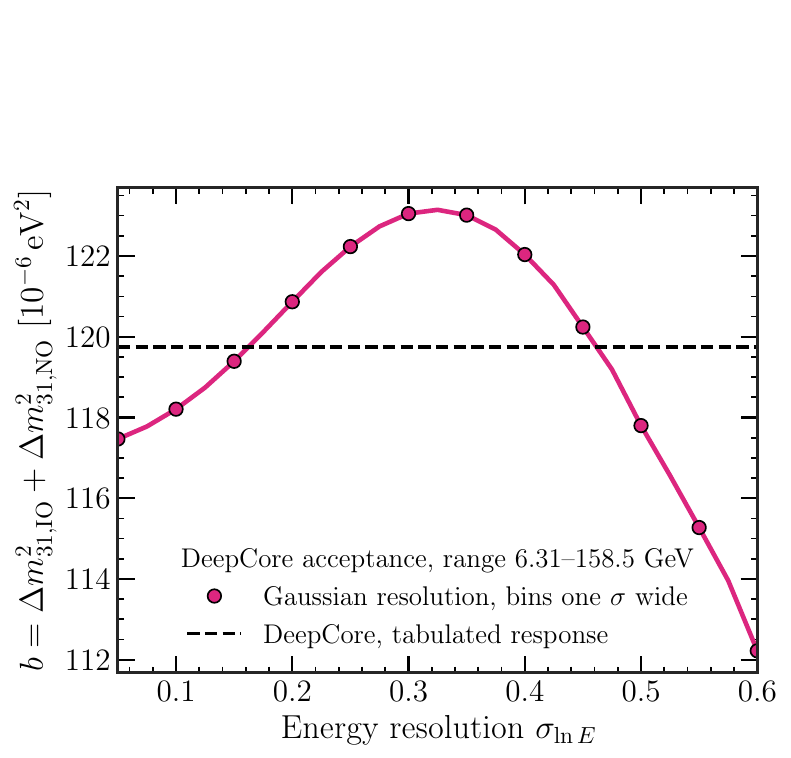}\hfill
\includegraphics[width=0.49\textwidth]{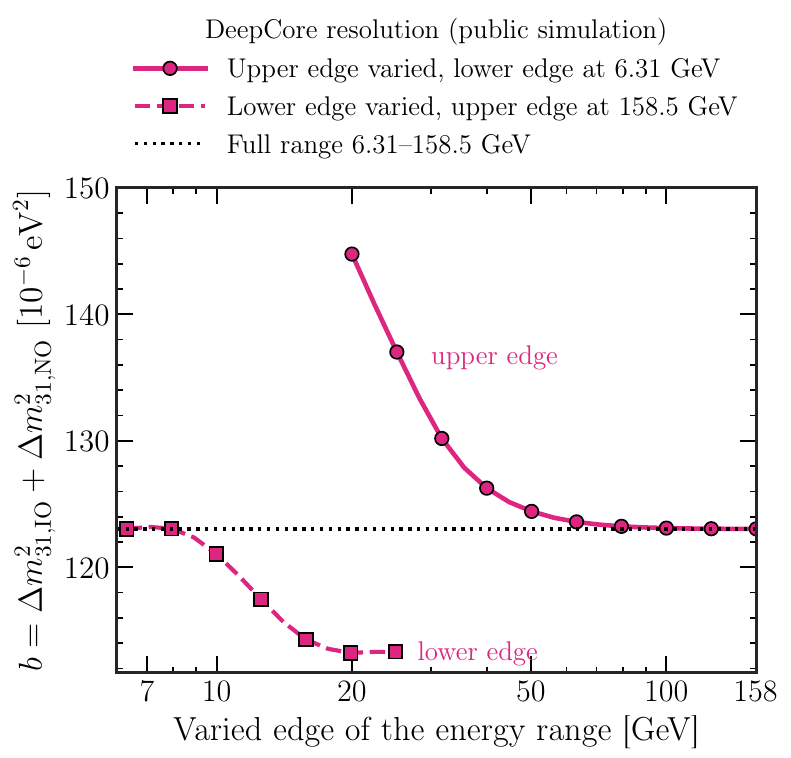}
\caption{Intercept of IceCube DeepCore using the acceptance of its public data release with Gaussian kernels in energy and zenith in place of the tabulated migration. Left: against the resolution $\sigma_{\ln E}$, with the zenith resolution extracted from the simulation of the public release, with energy bins one
standard deviation wide over the whole energy range published by DeepCore, 6.31--158.5~GeV. Dashed line: the tabulated response. Right: at the
energy resolution of DeepCore extracted from the simulation of its public release, $\sigma_{\ln E}\simeq0.29$--$0.34$, varying only the upper edge of the energy range used in the fit (circles, lower edge at 6.31~GeV) or only the lower
edge (squares, upper edge at 158.5~GeV). Dotted line: the full range, which differs from the dashed line of the left panel only by the Gaussian kernels in place of the tabulated migration. $\nu_\mu$ disappearance, oscillation parameters
from NuFIT~6.0, $\dmsq{31,\NO}=2.534\times10^{-3}\eVsq$.}
\label{fig:detector}
\end{figure}
Besides the oscillation parameters, the atmospheric solution depends on the detector response. The values $y_i$ and the weights $\omega_i$, and therefore $\datm$, depend on the energy and direction resolutions, which determine the oscillation phases averaged in each bin, and on the energy range. The ordering sensitivity of atmospheric detectors is known to depend on their energy and direction resolutions \cite{Akhmedov:2012ah,Winter:2013ema,Capozzi:2015bxa,IceCube:2023ins}, and the same holds for their wrong-ordering solution, by amounts small compared with its separation from a reactor experiment \cite{Blennow:2013vta}.
Our reference description of DeepCore is built from the simulated events of its public data release \cite{IceCube:B4RITM}. From these events we tabulate the probability that a neutrino of a given
flavor, energy, and direction is reconstructed in each analysis bin, and call this the tabulated response. Appendix~\ref{sec:s_valid} compares it with a full fit of the DeepCore public data, with all event classes and the systematic parameters of the release, and on Asimov data the two solutions agree within $2\times10^{-6}\eVsq$. To separate the roles of the resolution and of the energy range, we keep the DeepCore acceptance, that is, the expected event rate against the true energy and direction, and replace the migration from the true to the reconstructed variables with Gaussian kernels in energy and zenith.

Figure~\ref{fig:detector} shows the resulting intercept against the energy resolution $\sigma_{\ln E}$, the width of the Gaussian kernel in $\ln E$, which corresponds to a relative energy resolution $\sigma_E/E\simeq\sigma_{\ln E}$ at all energies (left, with the value of the tabulated response as a dashed line), and against the lower and the upper edge of the energy range (right, with the value for the full range as a dotted line). With energy bins that follow the resolution, the intercept changes by up to about $11\times10^{-6}\eVsq$ with the resolution, and not monotonically, because the resolution changes
both which oscillation phases each bin averages and which bins carry the largest weights. The energy range changes the intercept through the bins just above the MSW resonance, which have the largest $\delta_i$. Raising the lower edge of the range removes them and lowers the
intercept, while lowering the upper edge to a few tens of GeV increases their share of the weight and raises it.
A worse direction resolution, which \autoref{fig:detector} does not vary, lowers the intercept by less than the energy resolution moves it, as quantified in appendix~\ref{sec:s_tabresp}.

\subsection{Comparison of two atmospheric detectors}
\begin{figure}[t]\centering
\includegraphics[width=0.66\textwidth]{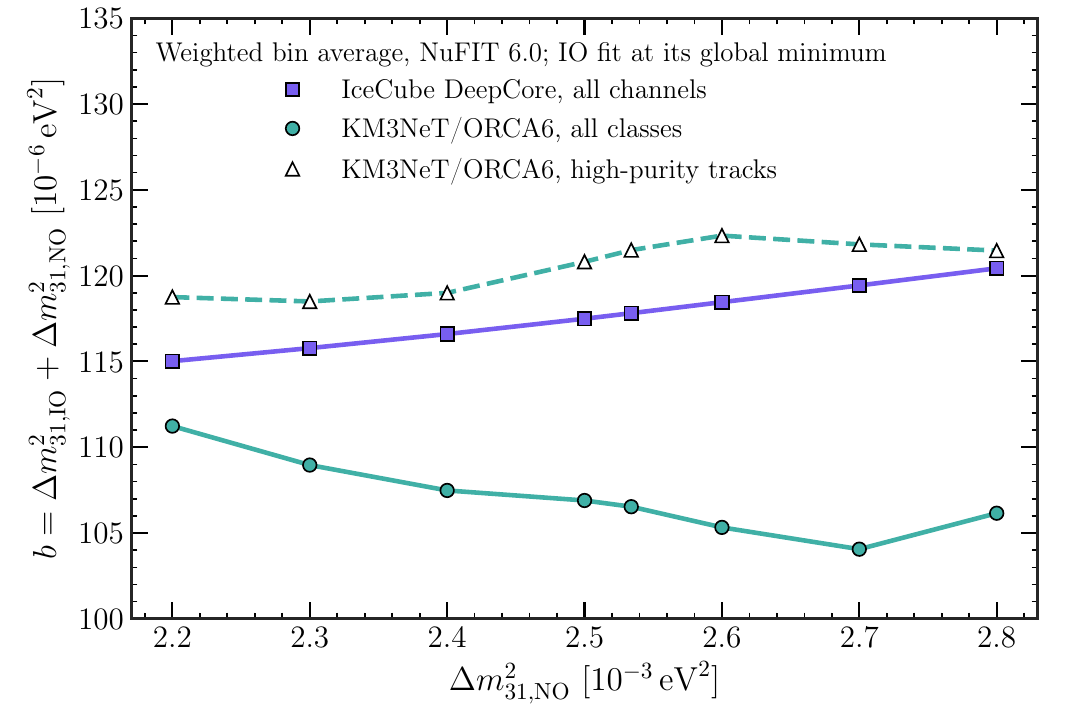}
\caption{Intercept of the wrong-ordering solution of IceCube DeepCore (tabulated response, all channels)
and KM3NeT/ORCA6 (public binned response \cite{KM3NeT:ORCA6data}, all classes and high-purity tracks) against the
true $\dmsq{31,\NO}$, from the weighted bin average with the IO fit at its global minimum in $\theta_{23}^{\IO}$ and $\dCP^{\IO}$. Other
oscillation parameters from NuFIT~6.0.}
\label{fig:orca}
\end{figure}
We then apply the weighted bin average to a second atmospheric detector, KM3NeT/\allowbreak ORCA \cite{KM3NeT:2021ozk}, whose atmospheric neutrinos have the same oscillation probabilities as those of DeepCore, so that only the detector differs. ORCA is located in sea water, its energy range starts at a lower energy, and it has its own energy and direction resolutions and three event classes (high-purity tracks, low-purity tracks, and
showers). The binned response of its first six detection units (ORCA6) is public
\cite{KM3NeT:ORCA6data,KM3NeT:ORCA6addendum,KM3NeT:2024flw}, and we evaluate the weighted bin average of \eqr{eq:master} on it with the
same oscillation probabilities.

Figure~\ref{fig:orca} shows the intercepts
of DeepCore with all channels, of ORCA6 with all event classes, and of the ORCA6 high-purity tracks alone, against the true
$\dmsq{31,\NO}$, with $\theta_{23}^{\IO}$ and $\dCP^{\IO}$ at the global minimum of the IO fit. At the
NuFIT~6.0 value of $\dmsq{31,\NO}$ the two detectors differ by about $11\times10^{-6}\eVsq$, and the high-purity tracks alone lie close to DeepCore. This difference comes mainly from the showers, which carry about $10\%$ of the ORCA6 weight and lower the intercept in two ways. They prefer bin intercepts far below those of the track bins, and they move the IO fit to a larger $\theta_{23}^{\IO}$. The intercepts of the two detectors change in opposite directions with the true splitting, and the slope of the ORCA6 solution stays within a few percent of $-1$. Two detectors of atmospheric neutrinos therefore have different wrong-ordering solutions, set by their energy and direction resolutions, their energy range, and their event classes.

In summary, Earth matter makes the energy and direction bins of atmospheric experiments prefer different values. The atmospheric solution therefore lies above the accelerator one, depends little on $\dCP$, moves with $\theta_{23}$ in $\nu_\mu$ disappearance, and depends on the energy and direction resolutions, the energy range, and the event classes of the detector.

\section{Wrong-ordering solutions of current experiments}
\label{sec:synergy}
\begin{figure}[t]\centering
\includegraphics[width=\textwidth]{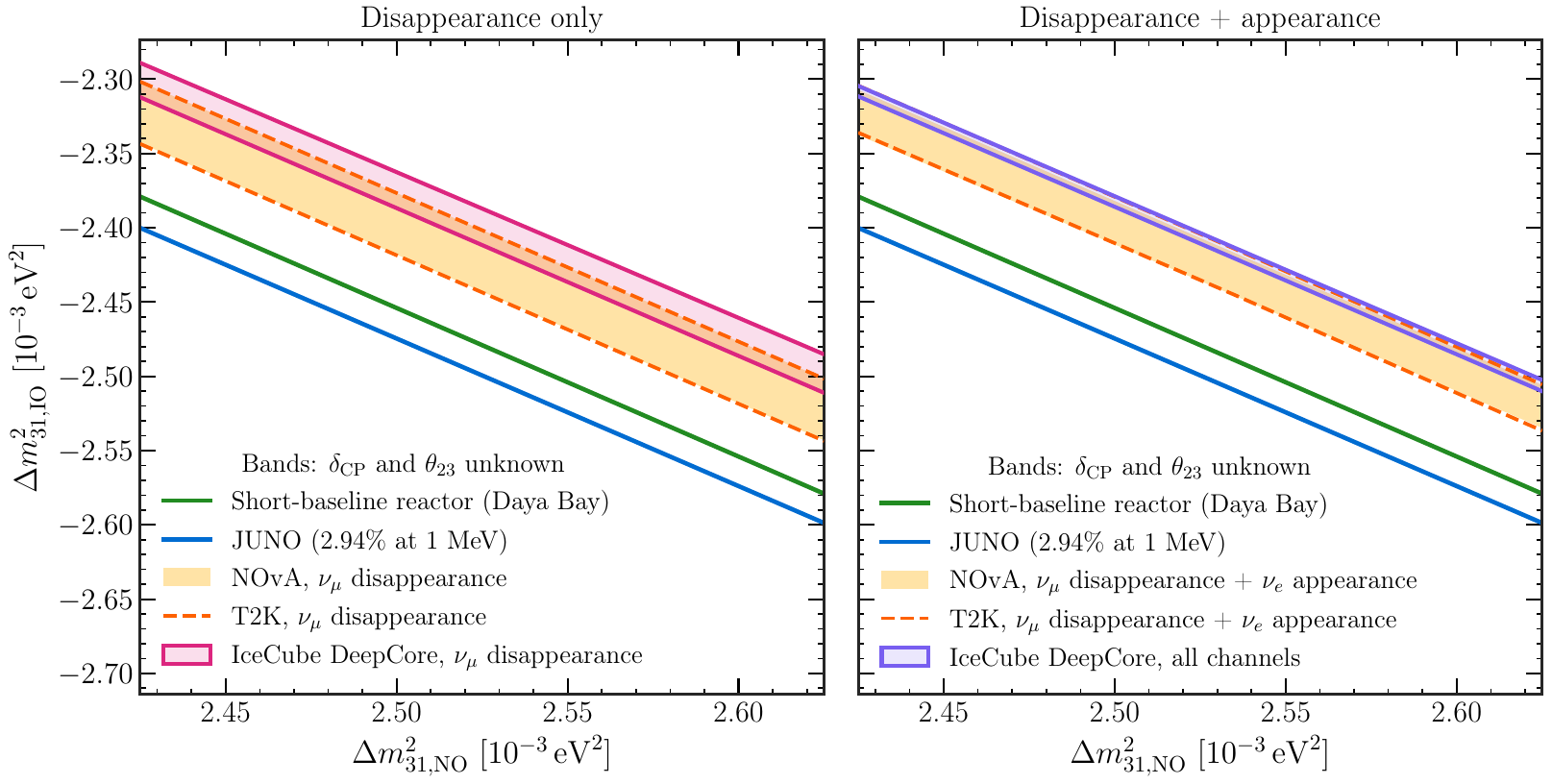}
\caption{Wrong-ordering solutions in the plane of $\dmsq{31,\IO}$ against $\dmsq{31,\NO}$, computed in this work
around the NuFIT~6.0 value of $\dmsq{31,\NO}$. Lines: Daya Bay (short-baseline reactor) and JUNO ($2.94\%$ at 1~MeV). Bands:
range of the solution over the unknown true $\dCP$ and $\theta_{23}$, for NOvA (filled), T2K (dashed lines), and
IceCube DeepCore (light fill with solid edges). Left: $\nu_\mu$ disappearance only. Right: with $\nu_e$ appearance for the
accelerators and all channels for DeepCore. The $3\sigma$ range of the true $\dCP$ covers nearly the whole circle, and that of $\sin^2\theta_{23}$ is $0.430$--$0.596$.}
\label{fig:plane}
\end{figure}

In this section we compare the wrong-ordering solutions of current experiments. We compute the solution of every experiment from \eqr{eq:master} as a function of the true $\dmsq{31,\NO}$. This gives the closed form of \eqr{eq:sbr} for the short-baseline reactor Daya Bay and the weighted bin average of \autoref{sec:vacuum} at its resolution of $2.94\%$ at 1~MeV for the medium-baseline reactor JUNO. For the accelerators NOvA and T2K, and for DeepCore, we evaluate the weighted bin average on their spectra or detector
responses. The accelerator spectra are built from public information \cite{NOvA:datarelease2024,NOvA:flux2017,T2K:flux2020}, with the exposures and backgrounds of the
experiments and statistical uncertainties only, as described in appendix~\ref{sec:s_lbl_real}.

\subsection{The plane of the two splittings}
Besides $\dmsq{31,\IO}$, the IO fit can adjust $\theta_{23}^{\IO}$ and $\dCP^{\IO}$. For every experiment we take the solution at the global minimum of the IO $\chi^2$ over $\dmsq{31,\IO}$, $\theta_{23}^{\IO}$ in both octants, and $\dCP^{\IO}$. Where the $\chi^2$ stays within $0.01$ of its minimum over a range of $\dCP^{\IO}$, we treat the minimum as degenerate and keep the whole range. In accelerator disappearance this range is the flat valley, which covers most or all of the circle \cite{Nunokawa:2005nx,Cabrera:2020ksc}. For DeepCore, also with all channels, it covers part or all of the circle. With $\nu_e$ appearance the accelerator minimum is instead a single point.

Figure~\ref{fig:plane} shows the solutions in the plane of $\dmsq{31,\IO}$ against $\dmsq{31,\NO}$. Daya Bay and JUNO
give one line each, since their solutions do not depend on $\dCP$ or $\theta_{23}$. The accelerators and DeepCore
give bands, whose width comes from the unknown true values of $\dCP$ and $\theta_{23}$ and, in accelerator disappearance,
also from the flat valley of $\dCP^{\IO}$. All lines and bands have
slope close to $-1$, and the bands of the accelerators and of DeepCore partly overlap. From the lowest to the highest intercept, the solutions are those of JUNO, Daya Bay, the accelerators, and DeepCore. JUNO lies below Daya Bay by the solar-phase term. The accelerators lie
higher, since $\kappa_\mu\simeq\cos^2\theta_{12}$ in $\nu_\mu$ disappearance, while $\kappa_e=\sin^2\theta_{12}$.
DeepCore lies higher still, mainly because of the matter shift contained in $\datm$.

\begin{figure}[t]\centering
\includegraphics[width=\textwidth]{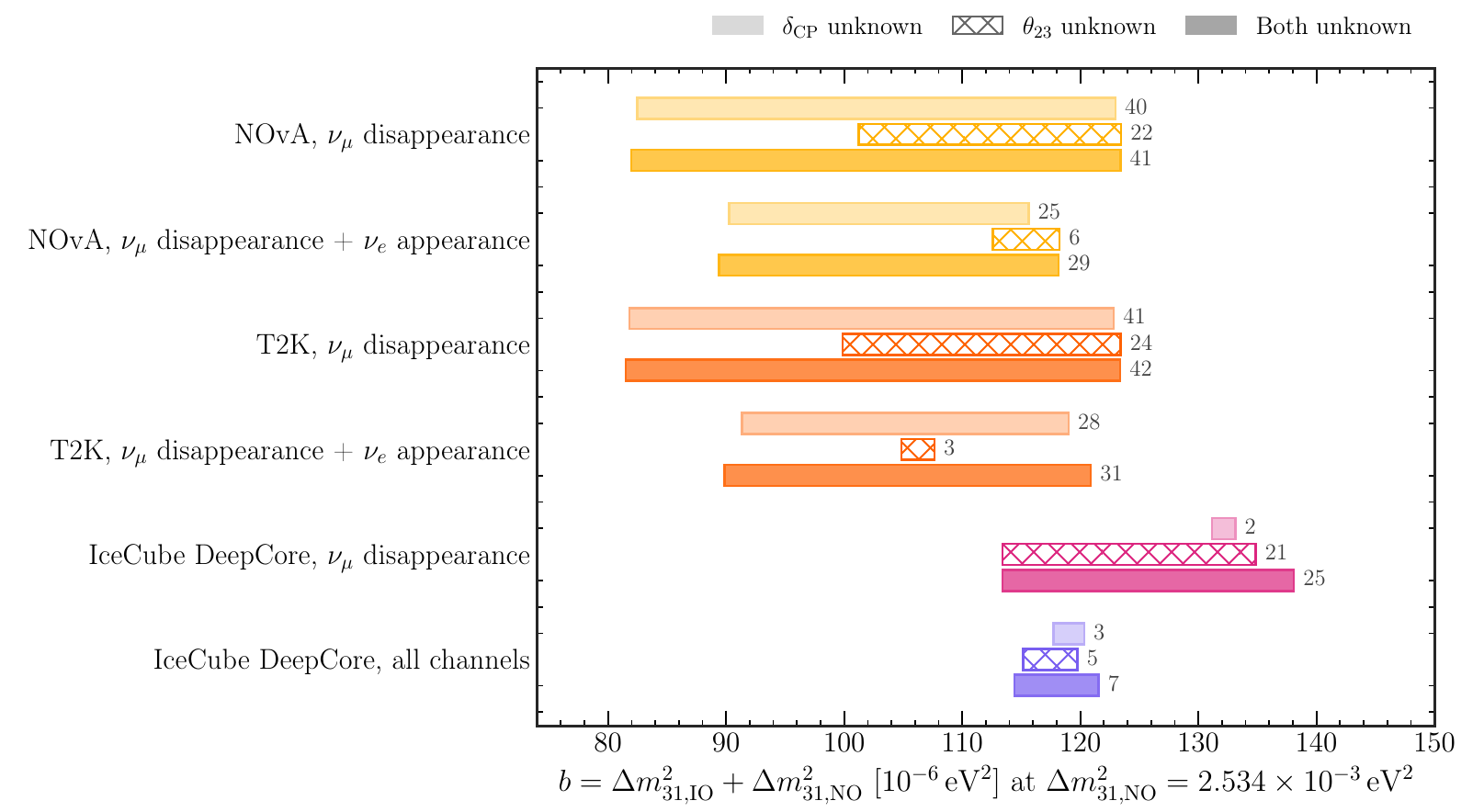}
\caption{Intercepts at $\dmsq{31,\NO}=2.534\times10^{-3}\eVsq$. For each sample, three bars give the range of $b$
when only the true $\dCP$ is varied over its NuFIT~6.0 $3\sigma$ range (true $\sin^2\theta_{23}$ at its NuFIT~6.0 value), when only the true $\theta_{23}$ is varied over its $3\sigma$ range
(true $\dCP$ at its NuFIT~6.0 value), and when both are varied. Each bar includes the range of $\dCP^{\IO}$ left by the IO fit, as the flat valley of an accelerator in disappearance. The widths in $10^{-6}\eVsq$ are given at the
right. The same IO fit, at its global minimum, is used in all three.}
\label{fig:bands}
\end{figure}

The true values of
$\dCP$ and $\theta_{23}$ are assumed unknown, and we vary them over their NuFIT~6.0 $3\sigma$ ranges for normal ordering
\cite{Esteban:2024eli}. The two ranges are varied independently, so the bands also contain combinations outside the joint $3\sigma$ region. The bands describe what is expected before the data are known. In nature $\dCP$ and $\theta_{23}$ have a single value, so each experiment has one solution inside its band. For accelerator disappearance alone this solution is itself an interval, because of the flat valley of $\dCP^{\IO}$.

\begin{figure}[t]\centering
\includegraphics[width=\textwidth]{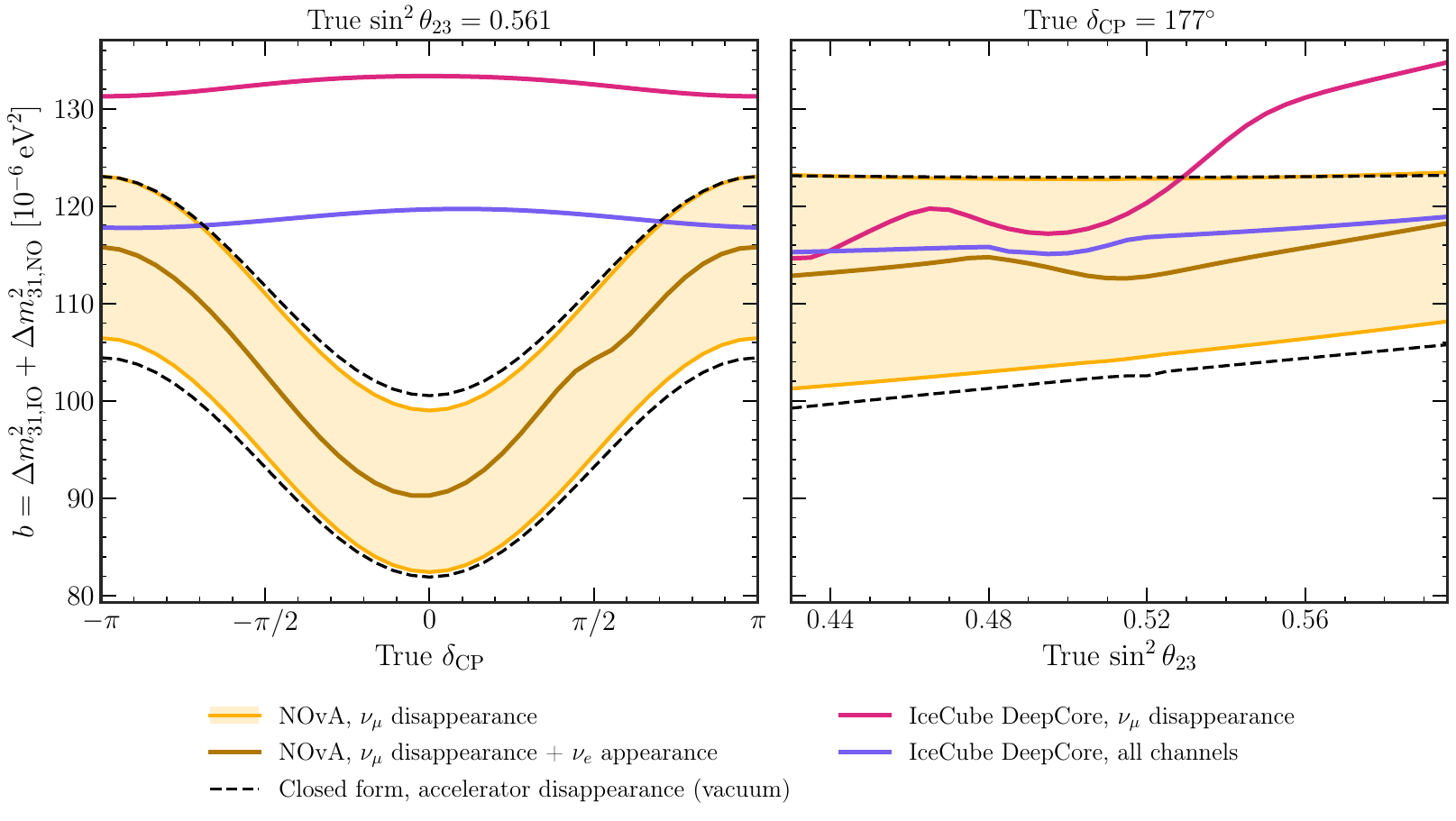}
\caption{Intercept at $\dmsq{31,\NO}=2.534\times10^{-3}\eVsq$ against the true $\dCP$ (left, true
$\sin^2\theta_{23}=0.561$) and the true $\sin^2\theta_{23}$ over its NuFIT~6.0 $3\sigma$ range (right,
true $\dCP=177^\circ$). The IO fit is at its global minimum in $\theta_{23}^{\IO}$ and $\dCP^{\IO}$.
NOvA disappearance (band): the flat valley of $\dCP^{\IO}$, where the $\chi^2$ is within $0.01$ of its minimum. Lines: NOvA with appearance (dark gold), DeepCore with $\nu_\mu$ disappearance (magenta) and with all channels
(violet), at the global minimum. For DeepCore the intercept changes little over the range of $\dCP^{\IO}$ within $0.01$ of the minimum, and this range is not drawn. Only NOvA is shown among the accelerators, and T2K has nearly the same disappearance band. All curves use statistical uncertainties only. Dashed: the accelerator closed form, \eqr{eq:lbl} in vacuum with the exact $\kappa_\mu$ of \eqr{eq:kmu_exact} \cite{Cabrera:2020ksc}, with $\dCP^{\IO}$ over its flat valley and $\theta_{23}^{\IO}$ the octant mirror of the true value, as returned by the fit to the NOvA spectrum.}
\label{fig:truth}
\end{figure}

\subsection{Ranges over \texorpdfstring{$\dCP$ and $\theta_{23}$}{delta CP and theta23}}
Figure~\ref{fig:bands} shows, for each experiment at the NuFIT~6.0 value of $\dmsq{31,\NO}$, three ranges of the
intercept: when only the true $\dCP$ is varied over its $3\sigma$ range, when only the true $\theta_{23}$ is varied,
and when both are varied. In disappearance all three bars of an accelerator experiment contain the flat valley of $\dCP^{\IO}$, which spreads the solution even when the true $\dCP$ and $\theta_{23}$ are known. The bar for an unknown $\theta_{23}$ is wide because of this valley, and $\theta_{23}$ itself adds little, while the unknown true $\dCP$ widens the bar further. The DeepCore $\nu_\mu$ solution is moved mainly by $\theta_{23}$, and with all channels
DeepCore is narrow in both parameters.

In disappearance, NOvA and T2K have nearly the same intercept, since $\kappa_\mu$ does not depend on the baseline, and their bands
differ by a few $10^{-6}\eVsq$. With appearance their bands are also close, but their solutions at the same true
values differ by up to about $20\times10^{-6}\eVsq$, mainly because the $\dCP^{\IO}$ returned by the IO fit of T2K, where the matter effect is smaller, changes with the true $\dCP$, while that of NOvA stays close to $270^\circ$.
Adding $\nu_e$ appearance to the IO fit constrains $\dCP^{\IO}$, but not the unknown true $\dCP$. The accelerator
bands in the right panel of \autoref{fig:plane} and in the rows of \autoref{fig:bands} with $\nu_e$ appearance are therefore
narrower, while the position of these bands is still determined mainly by disappearance. With appearance the IO fit cannot reproduce the NO data of disappearance and of appearance at once, and its minimum $\chi^2$ is larger than zero. This minimum is the intrinsic ordering sensitivity of the accelerator, the first term of \eqr{eq:dchi2}, and does not enter the synergy term.

In the following we show NOvA as an example. Its longer baseline gives a larger matter effect than at T2K, so it is the case where the full calculation differs most from the closed form of \eqr{eq:lbl}. Figure~\ref{fig:truth} shows how the intercepts of NOvA and DeepCore move with the true $\dCP$ (left) and the true $\sin^2\theta_{23}$
(right). For the accelerator the dashed curves of \autoref{fig:truth} show the closed form~\cite{Cabrera:2020ksc}, which follows from \eqr{eq:lbl} as
\begin{equation}
b_{\rm LBL}\simeq\Big[2c_{12}^2-s_{13}\sin2\theta_{12}\big(\tan\theta_{23}\cos\dCP+\tan\theta_{23}^{\IO}\cos\dCP^{\IO}\big)\Big]\dmsq{21},
\label{eq:lbl_band}
\end{equation}
where $\dCP$ and $\theta_{23}$ are the true values and $\dCP^{\IO}$ and $\theta_{23}^{\IO}$ those of the IO fit. The dashed curves use the exact $\kappa_\mu$ and, under the IO, the octant mirror of $\theta_{23}$ returned by the fit. In disappearance alone $\dCP^{\IO}$ spans the flat valley, and the closed form gives a band, whose center follows $\cos\dCP$ and whose width is set by the range of $\cos\dCP^{\IO}$ \cite{Nunokawa:2005nx,Cabrera:2020ksc}. With appearance the fit fixes $\dCP^{\IO}$, and the band reduces to a line.

The NOvA solution stays close to the closed form of \eqr{eq:lbl_band}. In disappearance the true $\theta_{23}$ and its octant mirror fit the NOvA data almost equally well. Matter effects change the effective $\theta_{13}$ in opposite directions for neutrinos and antineutrinos, and the IO fit returns the mirror, which describes both samples best. The same matter effects lower the effective $\theta_{13}$ of neutrinos under the IO, and with it the term in $\dCP^{\IO}$, so the NOvA band has nearly the same center as the closed form and is slightly narrower. With appearance, under the IO matter effects lower the appearance rate of neutrinos and raise that of antineutrinos, and the value $\dCP^{\IO}\simeq270^\circ$ returned by NOvA compensates this change most. Then $\cos\dCP^{\IO}\simeq0$, and the solution lies in the middle of the disappearance band. Against the true $\theta_{23}$ the accelerator band changes slowly, because $\theta_{23}$ enters it only through the small CP term, while the DeepCore $\nu_\mu$ disappearance solution moves with it, as described in \autoref{sec:atm_cp}. With all channels DeepCore stays nearly constant in both scans.

\begin{figure}[t]\centering
\includegraphics[width=\textwidth]{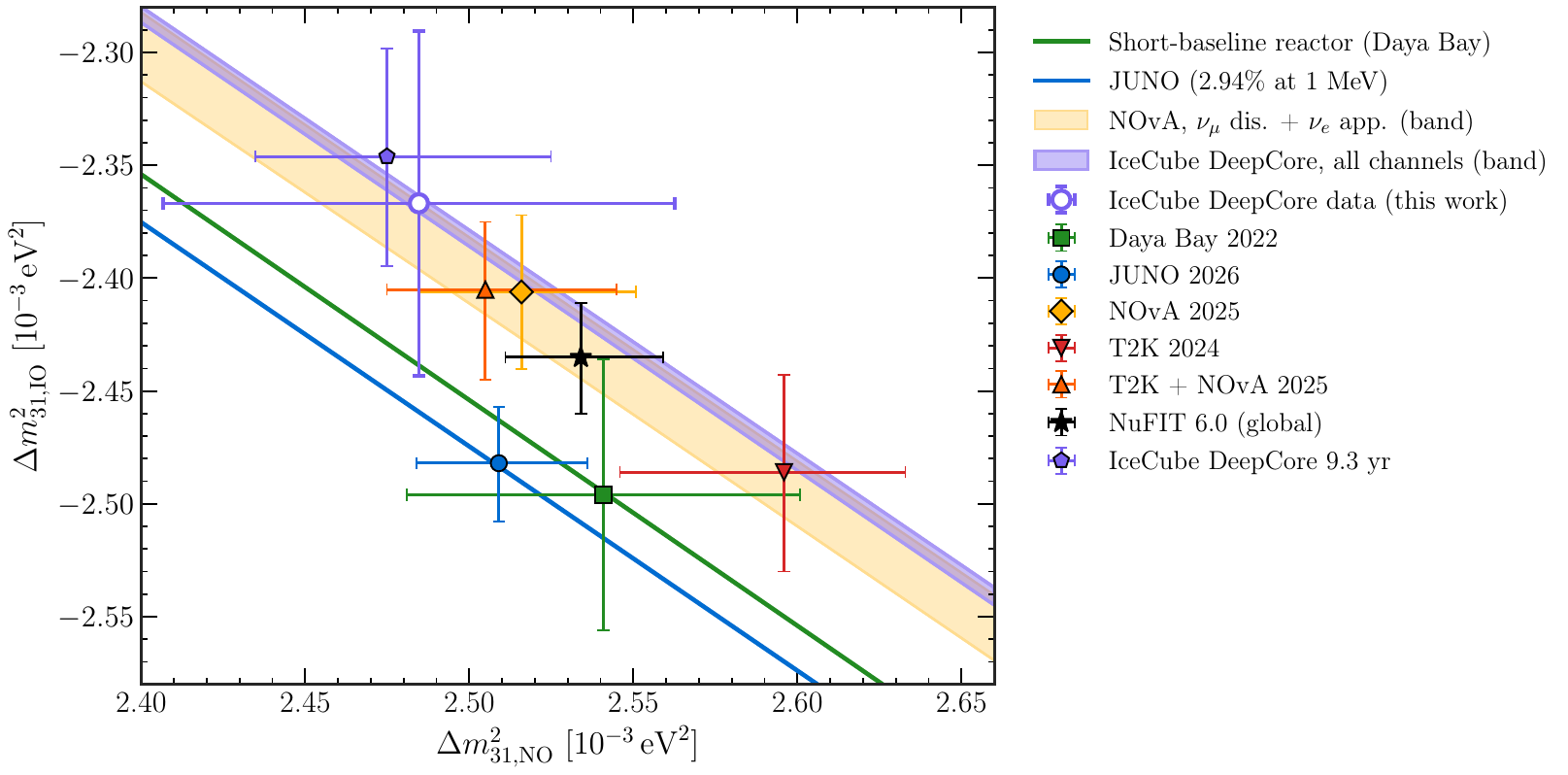}
\caption{Predicted wrong-ordering solutions (reactor lines and the bands of \autoref{fig:plane}, right: NOvA with appearance and IceCube
DeepCore with all channels) against the published best fits of $\dmsq{31}$ under the normal
and the inverted ordering, $1\sigma$: Daya Bay \cite{DayaBay:2022orm}, JUNO after 207 days
\cite{JUNO:2026osc}, NOvA \cite{NOvA:2025tmb} (frequentist best fit), the 2024 analysis of T2K \cite{T2K:HEPNP2025}, the joint NOvA and T2K analysis \cite{T2K:2025wet} (Bayesian, highest-posterior values), and
the NuFIT~6.0 global fit \cite{Esteban:2024eli}. Where $\dmsq{32}$ is quoted, we use $\dmsq{31}=\dmsq{32}+\dmsq{21}$ with the NuFIT~6.0 value of $\dmsq{21}$, and for IceCube DeepCore 9.3 years with the value used in its fit. Pentagon: IceCube DeepCore, 9.3 years \cite{IceCubeCollaboration:2024ssx}, with the
inverted-ordering best fit and interval taken from the released likelihood map \cite{IceCube:U20MMB},
profiled over $\sin^2\theta_{23}$. Open circle: our fit to the
IceCube DeepCore public data \cite{IceCube:B4RITM}, $1\sigma$, with the oscillation parameters of the release, as described in appendix~\ref{sec:s_iofit}.}
\label{fig:measured}
\end{figure}

\subsection{Comparison with published best fits}
Most experiments publish their best fits under both orderings. If the true ordering is normal, the best fit of an experiment under the IO is its wrong-ordering solution at its measured $\dmsq{31,\NO}$ and at the true $\dCP$ and $\theta_{23}$, up to statistical fluctuations, which move the two best fits nearly together along a line of slope close to $-1$. Each experiment therefore gives one point in the plane of \autoref{fig:plane}. Figure~\ref{fig:measured} shows the published best fits together with the lines and bands computed in
this work. Daya Bay \cite{DayaBay:2022orm} lies on its line. JUNO's measurement with 207 days of data \cite{JUNO:2026osc} gives $b=27\times10^{-6}\eVsq$, consistent with the line computed for the assumed resolution.\footnote{This measurement
uses a constraint from Daya Bay to select the physical minimum of the $\chi^2$, while the value and the precision of
$\dmsq{31}$ are set by JUNO.} NOvA \cite{NOvA:2025tmb}, T2K \cite{T2K:HEPNP2025}, and the joint NOvA and T2K analysis \cite{T2K:2025wet} lie inside the NOvA band with appearance,
and so does the global NuFIT~6.0 point \cite{Esteban:2024eli}, which precedes the JUNO measurement and whose $\dmsq{31}$ is dominated by the accelerators.

Our fit of both orderings to the DeepCore public data, described in appendix~\ref{sec:s_iofit}, lies inside the DeepCore all-channel band. Its position depends on how the detector corrections of the release, which are tabulated against the splitting of the normal-ordering simulation, are read in the IO fit. The 9.3-year DeepCore sample reconstructed with convolutional neural
networks \cite{IceCubeCollaboration:2024ssx}, whose likelihood is released in both orderings \cite{IceCube:U20MMB},
lies higher than our all-channel band. This sample has its own selection and reconstruction, with about seven times more events ($150\,257$ against $21\,402$), so its position is not a test of our modeling. The other published best fits lie on or inside the computed lines and bands. The published uncertainty of each splitting is larger than the distance between the accelerator and DeepCore bands, so these points cannot yet tell the two bands apart, and their agreement with the computed solutions is a check of the consistency of our inputs.

In summary, the published best fits agree with the computed solutions. The solution of JUNO lies far below those of the accelerators and of DeepCore, whose bands partly overlap. The position of the accelerator band is uncertain mainly through the unknown $\dCP$, and that of DeepCore through $\theta_{23}$ in $\nu_\mu$ disappearance, while with all channels DeepCore changes little.

\section{Mass ordering synergy in combinations}
\label{sec:comb}

\begin{figure}[t]\centering
\includegraphics[width=\textwidth]{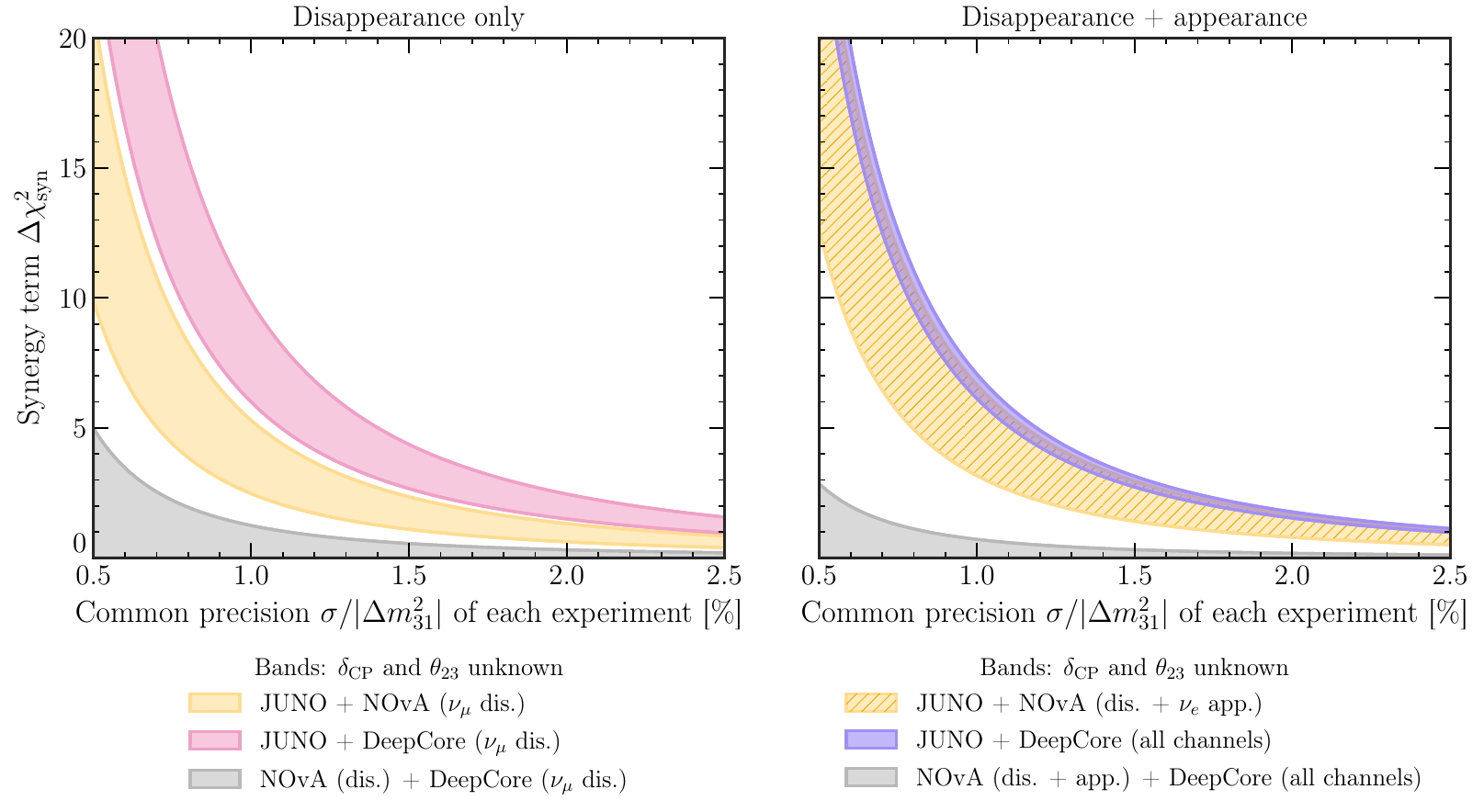}
\caption{The synergy term of \eqr{eq:dchi2} for pairs of JUNO ($2.94\%$ at 1~MeV), NOvA, and IceCube DeepCore, as a function of the relative precision $\sigma/|\dmsq{31}|$, common to the two experiments of a pair, at
$\dmsq{31,\NO}=2.534\times10^{-3}\eVsq$. Left: $\nu_\mu$ disappearance only for NOvA and DeepCore. Right: NOvA with $\nu_e$ appearance and
DeepCore with all channels. Bands: range over the unknown true $\dCP$ and $\theta_{23}$, the same for the two
experiments of a pair. Where the IO fit keeps a range of $\dCP^{\IO}$, the flat valley of NOvA in disappearance and the range of
DeepCore in both panels, the joint IO fit returns the value that brings the two solutions closest. In the right panel the JUNO--NOvA band is hatched. The precisions are illustrative. A joint fit of NOvA and DeepCore that shares $\theta_{23}^{\IO}$ and $\dCP^{\IO}$ gives a larger synergy, so the NOvA--DeepCore bands are lower estimates nearly everywhere.}
\label{fig:synergy}
\end{figure}

We now use the wrong-ordering solutions to evaluate the ordering sensitivity of a combination of two experiments. On NO Asimov data this sensitivity is the $\chi^2$ evaluated at the minimum of the joint IO fit. It contains the minima of the two individual IO fits, $\Delta\chi^2_A$ and $\Delta\chi^2_B$, and a term that arises because the two experiments return different values of $\dmsq{31,\IO}$. Under the NO hypothesis both experiments return the true splitting $x$, and under the IO hypothesis their own solutions $f_A(x)$ and $f_B(x)$. A joint IO fit has a single $\dmsq{31,\IO}$, which lies between them. Close to its minimum, the IO $\chi^2$ profile of each experiment is well approximated by a parabola in $\dmsq{31,\IO}$, with widths $\sigma_A$ and $\sigma_B$, respectively.\footnote{This holds
for $\dmsq{31}$ once the exposure is large enough. At low exposure the profile of a medium-baseline
reactor can have several local minima, as in the JUNO measurement with 207 days of data \cite{JUNO:2026osc}. } Minimizing the sum of the two parabolas gives \eqr{eq:dchi2}, where the approximate sign refers to the parabolic form of the profiles. Appendix~\ref{sec:s_comb} extends \eqr{eq:dchi2} to oscillation parameters shared by the two experiments and to real data, where the two experiments can measure different values of $\dmsq{31,\NO}$. The
synergy term of \eqr{eq:dchi2} depends only on the difference of the intercepts and on the precisions,
$(b_A-b_B)^2/(\sigma_A^2+\sigma_B^2)$. Figure~\ref{fig:synergy} shows it for the pairs formed by JUNO, NOvA, and DeepCore,
as a function of the relative precision $\sigma/|\dmsq{31}|$, with $|\dmsq{31}|$ the true NO splitting, common to the two experiments of a pair, with $\nu_\mu$
disappearance only (left) and with appearance and all channels (right). The true $\dCP$ and $\theta_{23}$ are
properties of nature and are the same for the two experiments of a pair, so each band is the range of the synergy
term over the unknown true values of $\dCP$ and $\theta_{23}$. In \autoref{fig:synergy} each experiment keeps its own $\theta_{23}^{\IO}$. Where its IO fit keeps a range of $\dCP^{\IO}$, as the flat valley of an accelerator in disappearance or the range of DeepCore, the joint IO fit takes the value that brings the two solutions closest, that is, the smallest $|b_A-b_B|$.

The intercept of JUNO lies far
below those of NOvA and DeepCore, so the pairs with JUNO give the largest synergy, in line with the large synergies reported in
reactor--accelerator and reactor--atmospheric studies
\cite{Nunokawa:2005nx,Cabrera:2020ksc,Parke:2024xre,Raikwal:2022nqk,NOvA:2026tir,Alves:2026ydc,Blennow:2013vta,IceCube-Gen2:2019fet,KM3NeT:2021rkn}.\footnote{The intercepts also depend on the solar parameters, which are fixed in this work. Within the precision of the JUNO measurement \cite{JUNO:2026osc}, they change the separations from JUNO by at most $2\times10^{-6}\eVsq$.}
The NOvA--DeepCore pair gives the smallest synergy, depending on the unknown $\dCP$ and $\theta_{23}$. For this pair \eqr{eq:dchi2} is an approximation, because the two experiments also share $\theta_{23}$ and $\dCP^{\IO}$. The synergy term of \eqr{eq:dchi2} treats the IO fits of the two experiments as independent. JUNO does not depend on $\theta_{23}$ and
$\dCP$, so in a pair with JUNO the two experiments share only $\dmsq{31,\IO}$, apart from $\theta_{13}$ and the solar parameters. \Eqr{eq:dchi2} then gives the joint result, provided that the solution and the precision of the partner are taken with $\theta_{23}^{\IO}$ and $\dCP^{\IO}$ profiled. NOvA and DeepCore both depend on $\theta_{23}$ and $\dCP$, and a joint IO fit shares these parameters between them. The synergy term then takes the form of \eqr{eq:pair_general}, derived in appendix~\ref{sec:s_comb}, where the two solutions differ also in $\theta_{23}^{\IO}$ and $\dCP^{\IO}$ and all differences are weighted by the inverse of the sum of the two covariance matrices. A joint fit of NOvA and DeepCore that shares $\theta_{23}^{\IO}$ and $\dCP^{\IO}$ gives a larger synergy at almost all true values of $\dCP$ and $\theta_{23}$, because the two IO fits prefer different values of these parameters also when they return the same octant, so the NOvA--DeepCore bands of \autoref{fig:synergy} are lower estimates nearly everywhere.

\begin{figure}[t]\centering
\includegraphics[width=\textwidth]{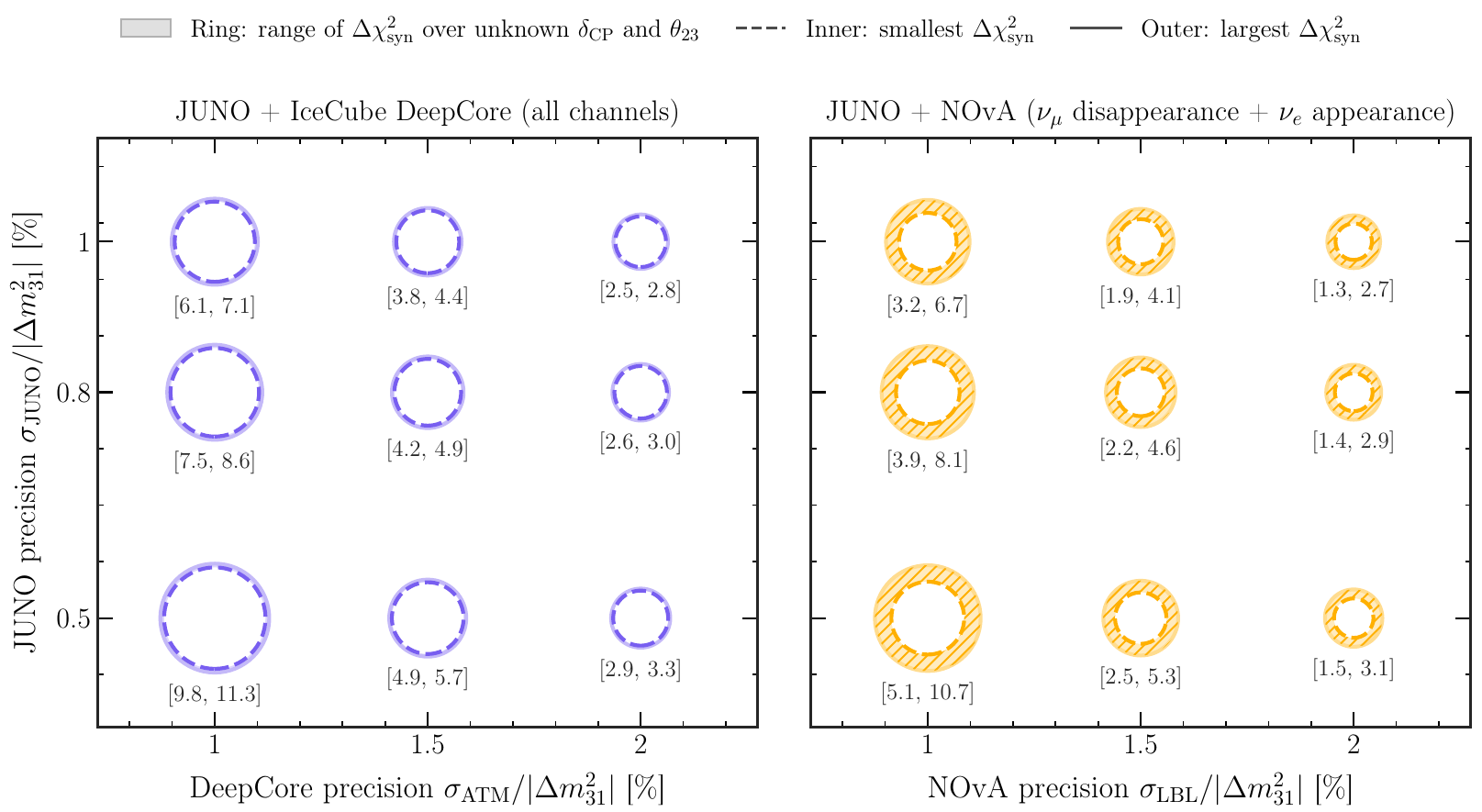}
\caption{Synergy term for JUNO with IceCube DeepCore (all channels, left) and with NOvA ($\nu_\mu$
disappearance and $\nu_e$ appearance, right), for different sets of precisions of the two experiments, relative to the true $|\dmsq{31,\NO}|$, with JUNO between $0.5\%$ and its present precision of $1.0\%$ \cite{JUNO:2026osc}. Each ring spans the range of $\Delta\chi^2_{\rm syn}$ over the unknown true $\dCP$ and
$\theta_{23}$ (inner dashed: smallest, outer: largest), with the joint IO fit as in \autoref{fig:synergy}. The area enclosed by each circle is proportional to $\Delta\chi^2_{\rm syn}$, and the rings of NOvA are hatched. The axis labels $\sigma_{\rm JUNO}$, $\sigma_{\rm ATM}$, and $\sigma_{\rm LBL}$ are the uncertainties $\sigma_A$ and $\sigma_B$ of \eqr{eq:dchi2} for JUNO, DeepCore, and NOvA. The precisions are illustrative.}
\label{fig:synergy_grid}
\end{figure}

Figure~\ref{fig:synergy_grid} shows the synergy term of JUNO with DeepCore (all channels, left) and with NOvA ($\nu_\mu$ disappearance and $\nu_e$ appearance, right) for several assumed precisions of the two experiments. For comparison, the present uncertainties on $|\dmsq{31}|$ or $|\dmsq{32}|$ are about $1.0\%$ for JUNO with 207 days of data \cite{JUNO:2026osc}, $1.4\%$ for NOvA \cite{NOvA:2025tmb}, and $1.9\%$ for DeepCore with 9.3 years of data \cite{IceCubeCollaboration:2024ssx}, a sample different from the one modeled here. Each ring corresponds to one pair of precisions, that of JUNO on
the vertical axis and that of the other experiment on the horizontal axis. Its inner and outer radii give the smallest and
the largest synergy over the unknown true $\dCP$ and $\theta_{23}$, and the two values are reported below the ring. The width of the rings shows how much the synergy depends on the unknown $\dCP$ and $\theta_{23}$.
For JUNO with DeepCore the ring is narrow. For a given detector response the synergy depends little on the unknown $\dCP$ and $\theta_{23}$, and it depends instead on the response, which has to be modeled for each sample. For JUNO with NOvA the ring is wide, and depending on the true $\dCP$ the same precisions give a synergy that changes by about a factor of two. At equal precisions the synergy with DeepCore is close to the largest value reached with
NOvA.

Combinations with JUNO already add to the ordering sensitivity in data. With constraints on $\dmsq{31}$ from long-baseline accelerator experiments, the first 207 days of JUNO data favor the normal ordering with a one-sided significance of at least $2.1\sigma$ for any value of $\dCP$ \cite{JUNO:2026osc}. Under the inverted ordering, the atmospheric data of Super-Kamiokande, IceCube DeepCore, and ORCA6 and the first 207 days of JUNO data prefer different values of $|\dmsq{31}|$, and their combination raises the preference for the normal ordering from $\Delta\chi^2=11.0$, which comes mainly from the intrinsic sensitivity of the atmospheric data, to $15.1$ \cite{Jin:2026evidence}.

In summary, the pairs with JUNO give the largest synergy. With an accelerator the synergy depends strongly on the unknown $\dCP$, and with an atmospheric experiment using all channels it is nearly fixed by the measured oscillation parameters and the detector response.

\section{Conclusions}
\label{sec:concl}
The sensitivity of a combination of experiments to the neutrino mass ordering depends on how far apart their
wrong-ordering solutions lie, that is, the values of $\dmsq{31,\IO}$ they return when NO data are fitted under the IO
hypothesis. We have shown how the wrong-ordering solution of any oscillation experiment is determined. It is the weighted average
of the values preferred by the individual bins of the measured event spectra, in reconstructed energy and, for atmospheric neutrinos, also zenith angle, with weights set by the statistical precision of each bin. Both the values and the weights depend on the detector response. In the plane of $\dmsq{31,\IO}$ against $\dmsq{31,\NO}$ every solution is a line of slope close to $-1$. The slope is exactly $-1$ for short-baseline reactors and accelerators in vacuum, and it differs from $-1$ by about $1\%$ for JUNO and DeepCore and by a few percent for ORCA6. Since the slopes differ from $-1$ by at most a few percent, each experiment is described by one number, its intercept
$b=\dmsq{31,\IO}+\dmsq{31,\NO}$, for atmospheric neutrinos in Earth matter as for reactors and accelerators, where
matter effects are small. All intercepts of this work are computed with the solar parameters and $\theta_{13}$ fixed and with statistical uncertainties only.

For short-baseline reactors, and for accelerators in vacuum, all bins prefer the same value, and the weighted bin average returns the known closed forms of $\dmsq{31,\IO}$ as a function of $\dmsq{31,\NO}$ \cite{deGouvea:2005hk,Nunokawa:2005nx,Minakata:2006gq,Minakata:2007tn,Parke:2016joa,Cabrera:2020ksc,Parke:2024xre}. The difference of two intercepts is the difference of the wrong-ordering splittings of the mass ordering sum rule \cite{Parke:2024xre}, so the weighted bin average extends the sum rule to any oscillation experiment. Systematic uncertainties act only on the weights of the bins, so to first order they cannot move the solution where all bins prefer the same value. At a medium baseline the solar phase varies across the spectrum, so the intercept of JUNO depends on its energy resolution. The accelerator solution depends on the unknown $\dCP$, and adding $\nu_e$ appearance narrows its range but leaves its position set by disappearance. For atmospheric neutrinos, Earth matter makes the bins prefer different values. The solution then depends on the energy and direction resolutions, the energy range, and the event classes of the detector, and it differs between DeepCore and ORCA6. Above the MSW resonance the frequency of $\nu_\mu$ disappearance is nearly independent of $\dCP$, so the atmospheric $\nu_\mu$ solution moves mainly with $\theta_{23}$.

The weighted bin average shows which features of an experiment set its wrong-ordering solution. The intercepts
of the individual experiments, computed in this way or read from their published best fits in the two orderings,
give the synergy of a combination before a joint fit, through \eqr{eq:dchi2} or, for measured values, through its form of appendix~\ref{sec:s_comb}. For combinations, the intercept $b$ quantifies the mismatch of the wrong-ordering solutions that earlier studies identified as the origin of the large synergies of reactors with accelerator and atmospheric experiments, and the intercept of JUNO lies far below those of both. Our results also show that the synergy of JUNO with an accelerator experiment depends strongly on the unknown $\dCP$, while with IceCube DeepCore using all channels it is nearly independent of $\dCP$ and $\theta_{23}$ but depends on the detector response, which therefore has to be modeled for each experiment.

As the precisions of JUNO and of the accelerator and atmospheric experiments improve, the separation of their intercepts sets how much their combination adds to the determination of the neutrino mass ordering.

\paragraph{Code and data availability.}
The code that reproduces all figures and numbers of this paper is public \cite{nmosynergies:2026}, and its documentation lists the version and the checksum of every input. All inputs are public: the IceCube DeepCore data releases \cite{IceCube:B4RITM,IceCube:U20MMB}, the KM3NeT/ORCA6 open-data release \cite{KM3NeT:ORCA6data}, the NOvA data and flux releases \cite{NOvA:datarelease2024,NOvA:flux2017}, the T2K flux release \cite{T2K:flux2020}, the GENIE cross sections \cite{Andreopoulos:2009rq,Nieves:2011pp,Schwehr:2016pvn}, the atmospheric neutrino fluxes of ref.~\cite{Honda:2015}, and the published configurations and results of the experiments cited in the text.

\acknowledgments
We thank our JUNO collaborators for useful discussions. We thank the IceCube, KM3NeT, NOvA, and T2K collaborations for making their data publicly available. During the preparation of this work the authors used AI tools to help edit the text, to assist with coding tasks,
and to cross-check numerical calculations. The
analysis, the physics content, and the style and content of the figures were defined by the authors, who verified
all derivations, numerical results, and statements and take full responsibility for the content of the paper.

\appendix
\addtocontents{toc}{\protect\setcounter{tocdepth}{1}}
\section{Inputs and experimental configurations}
\label{sec:s_setup}
\label{sec:s_experiments}
This appendix lists the oscillation parameters and describes how we model each experiment. These inputs
enter all results of sections~\ref{sec:formula} to~\ref{sec:comb}.

\subsection{Notation and oscillation parameters}
\label{sec:s_notation}
The true values of the oscillation parameters are the NuFIT~6.0 NO best fit without the Super-Kamiokande atmospheric data
\cite{Esteban:2024eli},
\begin{align}
s_{12}^2&=0.307,\quad s_{13}^2=0.02195,\quad s_{23}^2=0.561,\quad \dCP=177^\circ, \nonumber\\
\dmsq{21}&=7.49\times10^{-5}\eVsq,\quad \dmsq{31,\NO}=2.534\times10^{-3}\eVsq .
\label{eq:inputs}
\end{align}
We call the true $\dCP=177^\circ$ and the value of $\dCP^{\IO}$ given in \autoref{sec:formula} the NuFIT phases, and all results use them unless a range is shown. Two results use other values. The validation of appendix~\ref{sec:s_valid} uses the same $\dCP$ in both orderings, and the fits to the observed data of the DeepCore release in appendices~\ref{sec:s_repro} and~\ref{sec:s_iofit} use the fixed oscillation parameters of the release.
Intercepts are quoted in units of $10^{-6}\eVsq$. For reference, $2s_{12}^2\dmsq{21}=46.0\times10^{-6}\eVsq$ and
$2c_{12}^2\dmsq{21}=103.8\times10^{-6}\eVsq$.

\subsection{Reactor experiments}
\label{sec:s_reactors}
For the short-baseline reactor Daya Bay the solution is the closed form of \meq{eq:sbr}, which does not depend on the detector, so it needs no further input.

For the medium-baseline reactor JUNO we use its published configuration \cite{JUNO:2022osc}, with eight
cores, six of 2.9~GW$_{\rm th}$ at 52.1--52.8~km and two of 4.6~GW$_{\rm th}$ at 52.6--52.8~km, each
weighted by its thermal power over $L^2$. The flux is based on the Huber--Mueller model
\cite{Huber:2011wv,Mueller:2011nm}, with the fission fractions of a commercial pressurized-water reactor
($^{235}$U 0.58, $^{239}$Pu 0.30, $^{238}$U 0.07, $^{241}$Pu 0.05), and we use the inverse-beta-decay cross section \cite{Strumia:2003zx}. The bins span 1.81--9~MeV in reconstructed neutrino energy, in 20~keV steps, and the
energy resolution is evaluated at the prompt energy $E=E_\nu-0.78$~MeV,
\begin{equation}
\frac{\sigma_E}{E}=\sqrt{\Big(\frac{a_{\rm res}}{\sqrt{E/{\rm MeV}}}\Big)^2+b_{\rm res}^2+\Big(\frac{c_{\rm res}\,{\rm MeV}}{E}\Big)^2},
\end{equation}
where the stochastic term is $a_{\rm res}=2.61\%$, the constant term $b_{\rm res}=0.64\%$,
and the noise term $c_{\rm res}=1.20\%$, which give $\sigma_E/E=2.94\%$ at 1~MeV. These are the values predicted by the full simulation of the detector \cite{JUNO:2024reso} and used in its ordering sensitivity study \cite{JUNO:2024jaw}. With the unrounded terms the predicted resolution is $2.95\%$ at 1~MeV, and we use $2.94\%$ throughout. To vary the resolution we scale the three terms by the same factor, so that the energy dependence of the resolution stays that of the simulation and the resolution is perfect when the factor is zero. We label each resolution by its value at 1~MeV. The survival probability $P_{ee}$ is the exact
three-flavor one at constant density, computed with the NuFast algorithm \cite{Denton:2024pzc}. We use $\rho Y_e=1.225$~g\,cm$^{-3}$, with $Y_e=0.5$ and the average density $\rho=2.45$~g\,cm$^{-3}$ that JUNO estimates for the path of the antineutrinos through the crust and the sediment \cite{JUNO:2022osc}. This description contains statistical uncertainties only, without backgrounds. Within it the intercept is set
by the baselines, the reactor spectrum, and the energy resolution.

\subsection{Accelerator experiments}
\label{sec:s_lbl_real}
For all accelerators the oscillation probabilities are the exact three-flavor ones in matter of constant
crust density, computed with the NuFast algorithm \cite{Denton:2024pzc}, with the densities and the electron fraction listed with the other accelerator configurations in \autoref{tab:beams}.

The closed-form checks of \autoref{sec:vacuum} use a generic beam, with a cross section rising like $E$, events in 0.1~GeV bins over 0.5--5~GeV, and $\sigma_E/E=10\%$.

\begin{table}[t]\centering
{\small\begin{tabular}{@{}lll>{\raggedright\arraybackslash}p{0.25\textwidth}>{\raggedright\arraybackslash}p{0.20\textwidth}@{}}
\toprule
Configuration & $L$ & Density & Flux or spectrum & Exposure,\newline FHC and RHC\\
\midrule
Generic (\autoref{fig:vacuum}) & 810~km & $\rho=2.7$ & Gaussian at 2~GeV,\newline width 0.30~GeV & equal $\nu$ and $\bar\nu$ events\\
NOvA & 810~km & $\rho=2.84$ \cite{NOvA:2017abs} & 2024 data release \cite{NOvA:datarelease2024} & $26.6$ and $12.5\times10^{20}$ protons on target\\
T2K & 295~km & $\rho=2.6$ \cite{T2K:2021xwb} & 2020 flux release \cite{T2K:flux2020} & $14.9$ and $16.4\times10^{20}$ protons on target\\
\bottomrule
\end{tabular}}
\caption{Accelerator configurations, with densities in g\,cm$^{-3}$ and an electron fraction of 0.5. FHC and RHC are the forward and the reverse horn current, that is, the neutrino and the antineutrino mode.}
\label{tab:beams}
\end{table}

The realistic configurations of NOvA and T2K enter the bands of \autoref{sec:synergy}, with $\nu_\mu$ disappearance alone and together with $\nu_e$ appearance. We build their far-detector spectra from public information, separately for the neutrino and the antineutrino modes of the beam, with forward (FHC) and reverse (RHC) horn current. In $\nu_\mu$ disappearance the solution does not depend on the response, by \meq{eq:lbl}, up to the small matter shifts of appendix~\ref{sec:s_acc}.

Every sample is normalized to the published predicted event numbers, at the NOvA best fit for NOvA (table~II of ref.~\cite{NOvA:2025tmb}) and at $\dCP=-\pi/2$ for T2K (table~XI of ref.~\cite{T2K:2021xwb}). The cross sections therefore enter only through the ratio of the $\nu_\mu$ to $\bar\nu_\mu$ rates for NOvA and through the energy dependence of the rates for T2K. We use the cross sections of GENIE~2.12.10 \cite{Andreopoulos:2009rq} on water for T2K and on CH$_2$ for NOvA, built from its tabulated cross sections on oxygen, carbon, and hydrogen.

For NOvA the $\nu_\mu$-like samples are the no-oscillation prediction of the data release in reconstructed energy, split into $\nu_\mu$ and $\bar\nu_\mu$ by the ratio of their charged-current rates, each computed as the public near-detector flux \cite{NOvA:flux2017} times the cross section. The probability is averaged over each reconstructed-energy bin without energy migration. In the $\nu_e$-like samples the appeared $\nu_e$ are the no-oscillation $\nu_\mu$ spectrum times $P_{\mu e}$, the backgrounds have the shape of the no-oscillation $\nu_\mu$ spectrum, and the beam $\nu_e$ have the shape of the flux times the cross section.

For T2K we use the flux at Super-Kamiokande and one efficiency per signal or background component, which sets the relative size of the components in each sample. The efficiencies are calibrated to the event numbers of the Hyper-Kamiokande design report \cite{Hyper-Kamiokande:2018ofw}, which has the same beam line and detector technology, and the total of each sample is then normalized to the T2K prediction given above. The T2K spectra are folded with $\sigma_E/E=10\%$.

In both experiments the $\nu_\mu$-like count of a bin sums the surviving $\nu_\mu$ and $\bar\nu_\mu$, and the $\nu_e$-like count adds the appeared and the surviving beam $\nu_e$ and $\bar\nu_e$. The backgrounds are not oscillated. They are the neutral-current events and, for NOvA, the cosmic rays and the other backgrounds of its release.

\subsection{Atmospheric oscillation probabilities}
\label{sec:s_prop}
For DeepCore and ORCA6 we use the three-flavor probabilities through PREM of \autoref{sec:atm}, computed with the algorithm of NuFast-Earth \cite{Denton:2025nufastearth}. As in NuFast-Earth, each path is cut at the PREM discontinuities, and we divide each piece into layers of at most 500~km, each at its mean density. The eigenvalues and eigenvectors of each layer are obtained with the NuFast algorithm \cite{Denton:2024pzc}, and the amplitudes of the layers are multiplied along the path. Each path starts at a production height of 20~km in the atmosphere, treated as vacuum, and crosses the ice above DeepCore or the sea water above ORCA6.

\subsection{IceCube DeepCore}
\label{sec:s_tabresp}
We use the detector response tabulated from the public data release \cite{IceCube:B4RITM}, which contains the
high-purity ``golden'' event sample of ref.~\cite{IceCubeCollaboration:2023wtb}. The response is normalized
to the $21\,402$ neutrino events of this sample, the $21\,914$ observed events minus the 512 atmospheric muons expected from the simulation of the release. We use the azimuth-averaged Honda flux at the South Pole at solar minimum \cite{Honda:2015} and the probabilities described above.

The release provides the simulated events of the analysis. For each event it gives the flavor and charge, the true energy and direction, the analysis bin in which it is reconstructed, and a weight in GeV\,cm$^2$\,sr, which gives the
expected number of events when multiplied by the flux and the livetime. We use the events reconstructed
inside the published $10\times10$ analysis bins, nine uniform in $\ln E$ from 6.31 to 88.2~GeV and one from 88.2 to 158.5~GeV, and ten uniform in
$\cos\theta_z$ from $-1$ to $0.1$, with the event classes merged. From them we build three tables,
separately for each channel ($\nu_\mu$, $\nu_e$, and $\nu_\tau$ charged current, and neutral current by
flavor) and for neutrinos and antineutrinos. The acceptance $R(E,\cos\theta_z)$ is the summed event weight per unit $\ln E$ and $\cos\theta_z$, tabulated in true energy from 1 to 1000~GeV over the full sky. It
contains the cross section, the target mass, the trigger, and the selection. Each analysis bin $i=(i_E,i_C)$ is labeled by its reconstructed-energy bin
$i_E$ and its reconstructed-zenith bin $i_C$. The energy migration $K_E(E\to i_E)$ gives the fraction of events of a true-energy cell reconstructed in each analysis energy bin, and the zenith migration $K_C(E,\cos\theta_z\to i_C)$ the same for the analysis zenith bins. The migration tables are binned in coarse cells of true energy and zenith. They are filled with the event weight times the Honda flux of the event's own flavor, so that inside each coarse cell the migration follows the falling flux of the events. The expected count of the reconstructed bin $i=(i_E,i_C)$ is a sum over a finer grid of true cells $c$, and each fine cell uses the tables of the coarse cell that contains it,
\begin{equation}
\begin{split}
N_i\propto{}&\sum_{\nu,\bar\nu}\sum_k\sum_c R_{k,c}\,\Delta\ln E_c\,\Delta\!\cos\theta_{z,c}\;
\sum_\alpha\Phi_\alpha(E_c,\cos\theta_{z,c})\,P_{\alpha\beta}(E_c,\cos\theta_{z,c})\\
&\times K_{E,k}(E_c\to i_E)\,K_{C,k}(E_c,\cos\theta_{z,c}\to i_C),
\end{split}
\label{eq:Nb_tab}
\end{equation}
with the flux $\Phi_\alpha$ of each initial flavor $\alpha$ and the probability $P_{\alpha\beta}$ of
oscillation into the flavor $\beta$ of channel $k$ evaluated on the grid, with $R$, $K_E$, and $K_C$ the tables of channel $k$. The sum over $k$ runs over $\nu_\mu$ charged current alone or over all channels, and the counts are normalized to the events of the sample given above. The grid sizes are given in the public code \cite{nmosynergies:2026}, and a five times finer grid leaves the intercepts unchanged at the precision quoted in this paper. We make three approximations: the migration is factorized in energy and zenith, the event classes are merged, and the atmospheric-muon background is not included. Their combined effect on the intercept is evaluated in appendix~\ref{sec:s_cmp}. The release gives only the bin of each reconstructed event, not its reconstructed value, so the tabulated migration exists only for the published bins. Other bins and energy ranges, as in \autoref{fig:detector}, therefore need a model of the migration. For this model we extract the energy and zenith resolutions of the release from its $\nu_\mu$ charged-current events, as half the central 68\% interval of $\ln(E_{\rm reco}/E)$ and of $\cos\theta_{z,\rm reco}-\cos\theta_z$ in bins of true energy. As the reconstructed value of each event we take the center of its analysis bin, so the extracted widths also contain the spread of the events inside a bin, $w/\sqrt{12}$ for a bin of width $w$. The Gaussian kernels are integrated over the bin edges and add this spread again, so we subtract it in quadrature from the extracted widths.

For \autoref{fig:detector} we keep the tabulated acceptance and model the migration with Gaussian kernels in $\ln E$ and $\cos\theta_z$, integrated over the bin edges. In the left panel of \autoref{fig:detector}, which varies the energy resolution, the zenith resolution is the one extracted from the simulation of the release. In the right panel, which varies the energy range, the published bins are cut at the new edge, and the energy and zenith resolutions extracted from the simulation are used. With the extracted energy resolution and the bins of the release, scaling the zenith width of the Gaussian kernels from half to twice that of the release lowers the intercept by $6.5\times10^{-6}\eVsq$ for $\nu_\mu$ disappearance and by $5.0\times10^{-6}\eVsq$ with all channels. At the resolutions of the release and over the full energy range, the Gaussian kernels and the tabulated response, two models of the same detector, give intercepts $3\times10^{-6}\eVsq$ apart.

\subsection{KM3NeT/ORCA6}
\label{sec:s_orca}
The ORCA6 open-data release \cite{KM3NeT:ORCA6data,KM3NeT:ORCA6addendum} (433~kt\,yr, the sample of
ref.~\cite{KM3NeT:2024flw}) gives the expected weight of each true cell in each reconstructed bin, for every
neutrino flavor, charge, and interaction. The reconstructed bins are 10 up-going $\cos\theta_z$ bins in each of three classes, namely high-purity tracks, low-purity tracks, and showers. In energy the two track classes have 14 bins between 2 and 100~GeV, and the showers have one more bin, from 100 to 1000~GeV \cite{KM3NeT:2024flw}. The count of a reconstructed bin is the sum of these weights times the oscillated flux of the
initial flavors, plus the atmospheric-muon background of the release. We use the Honda flux
\cite{Honda:2015} at the Fr\'ejus site (solar minimum, no mountain), as prescribed by the release
\cite{KM3NeT:ORCA6data}, and the probabilities of appendix~\ref{sec:s_prop}, averaged over each true cell on a finer grid. Without any nuisance parameter
the predicted total agrees with the observed one to $0.7\%$. The high-purity tracks, the low-purity tracks, and the showers carry $65\%$, $25\%$, and $10\%$ of the weight, respectively. Adding the low-purity tracks to the high-purity ones changes the intercept by less than $2\times10^{-6}\eVsq$. Adding the showers lowers it by $13\times10^{-6}\eVsq$, and the fitted $\sin^2\theta_{23}$ rises by about $0.01$. Since ORCA6 with all classes lies about $11\times10^{-6}\eVsq$ below DeepCore, as shown in \autoref{fig:orca}, the high-purity tracks alone lie within a few $\times10^{-6}\eVsq$ of DeepCore.

\section{Derivation of the weighted bin average}
\label{sec:s_formula}
This appendix derives the weighted bin average of \autoref{sec:formula}, its symmetry under the exchange
of the orderings, its iterative solution, and its form with profiled nuisance parameters.

\subsection{Bin observable, Poisson weight, and stationarity}
\label{sec:s_stat}
The expected count in a bin $i$ of reconstructed variables, under ordering $H$
and for a given value of $\dmsq{31}$, is
\begin{equation}
N_i^H=\int\!dE\int\!d\cos\theta_z\;G_i(E,\cos\theta_z)\,P^H(E,\cos\theta_z),
\qquad G_i=\Phi\,\sigma\,\varepsilon\,T\,K_i ,
\label{eq:Nb}
\end{equation}
where the integrals run over the true energy and direction, $\Phi$ is the flux,
$\sigma$ the cross section, $\varepsilon$ the efficiency, $T$ the exposure, and
$K_i$ the probability that an event at the true point is reconstructed in bin
$i$. Only $P$ depends on the splitting, so the count factorizes exactly,
$N_i^H=N^{\rm u}_iP_i^H$, with the unoscillated count $N^{\rm u}_i=\int\!dE\int\!d\cos\theta_z\,G_i(E,\cos\theta_z)$. The bin observable
$P_i^H$ is the probability averaged over the true energies and directions that the bin collects. A bin that collects several channels has no single probability, so we define $P_i^H=N_i^H/N_i^{\rm ref}$ with any fixed reference count $N_i^{\rm ref}$. The choice of $N_i^{\rm ref}$ does not matter, since neither $y_i$ nor $\omega_i=(\partial_yN_i^{\IO})^2/N_i^{\NO}$ depends on it. For Asimov data the $\chi^2$ of \meq{eq:chi2} is twice the Poisson log-likelihood ratio, expanded to second order in $N_i^{\IO}-N_i^{\NO}$, with $W_i$ the inverse variance of the measured fraction $P_i$.

Expanding $\PIO$ around a point $y$ gives the Newton step
\begin{equation}
y_i\simeq y+\frac{\PNO-\PIO(y)}{s_i(y)},
\label{eq:newton}
\end{equation}
with $s_i=\partial_y\PIO$. The Newton step approximates the exact root of $\PIO(y_i)=\PNO$, where one exists, with an error of order $(y_i-y)^2\partial_y^2\PIO/s_i$. Taken at $y=f$ as an equality, the Newton step is the definition of $y_i$ in \meq{eq:yi}. The minimum of \meq{eq:chi2} satisfies
\begin{equation}
\sum_i W_i\,s_i(f)\,\big[\PNO-\PIO(f)\big]=0 .
\label{eq:stat}
\end{equation}
By the definition of $y_i$, $\PNO-\PIO(f)=s_i(y_i-f)$, and \eqr{eq:stat} becomes
$\sum_i W_i s_i^2(y_i-f)=0$, which is \meq{eq:master}. The relation is exact for the $\chi^2$ of
\meq{eq:chi2}.

\subsection{Exchanging the two orderings}
\label{sec:s_exchange}
For a true inverted ordering with splitting $y$, the fit under the NO hypothesis returns $x'$. Each bin prefers the value $x'_i$ given by \eqr{eq:newton} with the two orderings exchanged, including their values of $\theta_{23}$ and of the CP phase. Where all bins agree, as for short-baseline reactors and accelerators in
vacuum, a true inverted ordering gives $x'=-y+b$ with the same intercept. In matter the two orderings differ in amplitude as well as in phase. For a true inverted ordering both the bin intercepts and the weights then change, and the two changes partly cancel. We check this for DeepCore with $\theta_{23}$ fixed. We take the IO solution $f(x)$ as the true splitting of an inverted ordering and fit it under the NO hypothesis. The fit returns the original $x$ to about $2\times10^{-6}\eVsq$ for $\nu_\mu$ disappearance and to within $0.1\times10^{-6}\eVsq$ with all channels, so the intercept is nearly the same for the two true orderings. The synergy term of \meq{eq:dchi2} is therefore symmetric under the exchange of the two orderings when, besides the intercepts, the widths $\sigma_A$ and $\sigma_B$ of the wrong-ordering fits are also the same for the two true orderings.

\subsection{Gauss--Newton iteration}
\label{sec:s_gn}
Inserting \eqr{eq:newton} at a trial value $y_k$ into \meq{eq:master} gives the Gauss--Newton step for the $\chi^2$ of \meq{eq:chi2}, whose fixed point satisfies \eqr{eq:stat}. For the atmospheric evaluation we use the same step written in counts, with
$\partial_yN_i^{\IO}$ in place of $s_i$ and $1/N_i^{\NO}$ in place of $W_i$. The iteration starts from the closed forms of \autoref{sec:vacuum}. All intercepts with fixed $\theta_{23}$ and $\dCP^{\IO}$ are obtained in this way. The bands, which
take the global minimum over $\theta_{23}$ and $\dCP^{\IO}$, use the same $\chi^2$, minimized as follows. As shown in appendix~\ref{sec:s_cp}, the Hamiltonian in matter depends on $\theta_{23}$ and $\dCP$ only through a rotation
of the flavor basis, so at fixed $\dmsq{31,\IO}$ every expected count is a trigonometric polynomial of degree two
in $2\theta_{23}$ and in $\dCP^{\IO}$, for any matter profile \cite{Yokomakura:2002av,Kimura:2002wd}. A polynomial of degree two in an angle has five terms, a constant and the cosines and sines of the angle and of twice the angle, so 25 coefficients describe each count. We compute them once and minimize the $\chi^2$ in $\dmsq{31,\IO}$, in $\sin^2\theta_{23}$ separately
in each octant, and in $\dCP^{\IO}$. The grids, the tolerances, and the sampling of the true $\dCP$ and $\theta_{23}$ are given in the public code \cite{nmosynergies:2026}.

\subsection{Profiled nuisance parameters}
\label{sec:s_nuis}
The validation of appendix~\ref{sec:s_valid} compares the weighted bin average with a full fit that profiles the
systematic uncertainties. This requires the weighted bin average with profiled nuisance parameters, which we
derive here. Let the IO fit also vary nuisance parameters $\eta$, with Jacobian
$J_\eta=\partial N/\partial\eta$, constrained by pull terms~\cite{Fogli:2002pt} with prior covariance $\Sigma$. A parameter
without a prior, such as $\theta_{23}$, has a vanishing row and column in $\Sigma^{-1}$. From here on we work in counts, as in the Gauss--Newton step of appendix~\ref{sec:s_gn}. We write $s_i=\partial_yN_i^{\IO}$, which is $N^{\rm u}_i$ times the derivative of \autoref{sec:formula}, and we redefine $W_i=1/N_i^{\NO}$, the weight of \autoref{sec:formula} divided by $(N^{\rm u}_i)^2$, so that $\omega_i=W_is_i^2$ keeps its value. With the residual $r=N^{\NO}-N^{\IO}$ and the diagonal matrix $W$ of these weights, the linearized $\chi^2$ is
\begin{equation}
\chi^2=(r-s\,\Delta y-J_\eta\Delta\eta)^{\rm T}W(r-s\,\Delta y-J_\eta\Delta\eta)+\Delta\eta^{\rm T}\Sigma^{-1}\Delta\eta .
\end{equation}
Minimizing over $\Delta\eta$ leaves $\chi^2=(r-s\,\Delta y)^{\rm T}\tilde W(r-s\,\Delta y)$, with
$\tilde W=W-WJ_\eta(J_\eta^{\rm T}WJ_\eta+\Sigma^{-1})^{-1}J_\eta^{\rm T}W$. This matrix satisfies
$\tilde Ws=W\tilde s$, with the profiled derivative
\begin{equation}
\tilde{s}=s-J_\eta\big(J_\eta^{\rm T}WJ_\eta+\Sigma^{-1}\big)^{-1}J_\eta^{\rm T}W\,s,
\label{eq:stilde}
\end{equation}
and $r_i=s_i(y_i-y)$ by the Newton step of \eqr{eq:newton} taken as an equality. The minimum in $y$ is therefore at
\begin{equation}
f=\frac{\sum_i W_i\,s_i\,\tilde s_i\,y_i}{\sum_i W_i\,s_i\,\tilde s_i},
\label{eq:master_nuis}
\end{equation}
the weighted bin average with the weights $\tilde\omega_i=W_i s_i\tilde s_i$. The profiled derivative
$\tilde s$ is the part of $s$ that no combination of nuisance parameters can reproduce within their priors.
A bin whose derivative with respect to the splitting is mimicked by a nuisance parameter loses weight, and the sum of the
weights is the inverse variance of the splitting with the nuisance parameters profiled. The sum,
$s^{\rm T}\tilde Ws$, is positive, while a single $\tilde\omega_i$ can be negative in bins that a nuisance parameter
correlates strongly with others, so the $\tilde\omega_i$ are the weights of a linear estimator rather than
probabilities. For nuisance parameters that stay close to their prior centers the $y_i$ do not change and only the weights do. The fitted values of $\theta_{23}^{\IO}$ and $\dCP^{\IO}$, in contrast, can move far from the true ones, as \autoref{sec:atm_cp} and \autoref{sec:synergy} show. We therefore evaluate $P_i^{\IO}$, the $y_i$, and the $\omega_i$ at these fitted values. The $\chi^2$ is stationary in $y$ at the minimum, so \meq{eq:master} holds there exactly with $\omega_i=W_is_i^2$.

\section{Disappearance in vacuum: reactor and accelerator experiments}
\label{sec:s_vacuum}
This appendix gives the effective-frequency form of the survival probability used in \autoref{sec:vacuum},
the matter effects in accelerator disappearance, and the dependence of the JUNO solution on the energy
resolution.

\subsection{Effective-frequency form of the survival probability}
\label{sec:s_efffreq}
In vacuum, with $\Delta_{ij}=\dmsq{ij}L/4E$ and $u_k=|U_{\alpha k}|^2$, the two fast terms of the survival
probability combine into one oscillation, \meq{eq:twofreq_main}
\cite{Nunokawa:2005nx,Minakata:2006gq,Minakata:2007tn,Parke:2016joa}, where $\Delta_{\alpha\alpha}=\dmaa L/4E$ is the phase of the effective splitting of \meq{eq:dmaa}. The amplitude and the solar phase are
\begin{equation}
C^2_{\alpha\alpha}=1-\frac{4u_1u_2\sin^2\Delta_{21}}{(u_1+u_2)^2},\qquad
\tan\varphi_{\alpha\alpha}=\frac{u_1\sin\frac{2u_2\Delta_{21}}{u_1+u_2}-u_2\sin\frac{2u_1\Delta_{21}}{u_1+u_2}}
{u_1\cos\frac{2u_2\Delta_{21}}{u_1+u_2}+u_2\cos\frac{2u_1\Delta_{21}}{u_1+u_2}} .
\end{equation}
The second expression is equivalent to \meq{eq:phi_main}. For equal mixing parameters in the two orderings, the coefficients $u_k$ and $\Delta_{21}>0$ are the same, so the ordering enters only
through the sign in front of $\varphi_{\alpha\alpha}$. When the solar oscillation has not developed, $C_{\alpha\alpha}$ is close to one. Matching the two cosines at one energy gives \meq{eq:yb_vac}, with $\kappa_\alpha=u_2/(u_1+u_2)$, $|\dmsq{\alpha\alpha}|^{\NO}=x-\kappa^{\NO}_\alpha\dmsq{21}$, and $|\dmsq{\alpha\alpha}|^{\IO}=-y+\kappa^{\IO}_\alpha\dmsq{21}$. This phase-matching value equals $y_i$ of \eqr{eq:yi} under two conditions. First, $P_i^{\IO}$ is close to linear between $f$ and $y_i$. Second, for $\alpha=\mu$, the two orderings have the same CP phase. With different phases the terms of $P_{\mu\mu}$ that oscillate with $\dmsq{21}$ differ between the orderings and add to $y_i$ an offset of order $(\dmsq{21})^2/|\dmsq{31}|$, below $1\times10^{-6}\eVsq$ for DeepCore without matter. For $\alpha=e$ at short baselines $\varphi_{ee}=\mathcal{O}(\Delta_{21}^3)$~\cite{Parke:2016joa}. At 4~MeV and 1.65~km, the flux-averaged baseline of the far hall of Daya Bay \cite{DayaBay:2022orm}, $\Delta_{21}\simeq0.04$, so $\varphi_{ee}$ is negligible and the matching gives \meq{eq:sbr}.

\subsection{Accelerator \texorpdfstring{$\nu_\mu$}{numu} disappearance and matter effects}
\label{sec:s_acc}
For accelerators the solar phase is negligible, since $\Delta_{21}\simeq0.04$ at 810~km and 2~GeV. For $\alpha=\mu$ in vacuum \meq{eq:master} returns \meq{eq:lbl} for any binning and resolution.
Matter makes the bins differ only slightly, in line with the small size of matter effects in long-baseline
$\nu_\mu$ disappearance \cite{Denton:2024thm}, and the intercept depends on the share of antineutrino events.
The matter effects shift the solution in opposite directions for neutrinos and antineutrinos, so the shift cancels for equal shares of the two, as in the generic beam of \autoref{fig:vacuum}, and each experiment is evaluated with its own share. For a beam of neutrinos alone, with $\theta_{23}$ fixed at the same value in both orderings, the intercept at 810~km lies $4$ to $7\times10^{-6}\eVsq$ below the closed form with the flux of the generic beam, depending on the true $\dCP$. At 295~km, with a flux peaked at 0.6~GeV, it lies less than $0.5\times10^{-6}\eVsq$ away from the closed form. The same matter effects change the width of the flat valley. Under the IO they reduce the effective $\sin2\theta_{13}$ of neutrinos, roughly as $1/(1+\hat A)$ with $\hat A=2\sqrt2G_FN_eE/x\simeq0.17$ at 2~GeV, and increase it for antineutrinos. The term of $\kappa_\mu$ that depends on $\dCP^{\IO}$ is proportional to $\sin\theta_{13}$, as \meq{eq:lbl} shows, so it is reduced for neutrinos. The NOvA sample consists mostly of neutrinos and therefore depends less on $\dCP^{\IO}$. Its disappearance band in \autoref{fig:truth} is about $11\%$ narrower than that of the closed form, with its center close to that of the closed form.

\subsection{Medium-baseline reactor: solar phase and energy resolution}
\label{sec:s_mbr}
\begin{figure}[t]\centering
\includegraphics[width=\textwidth]{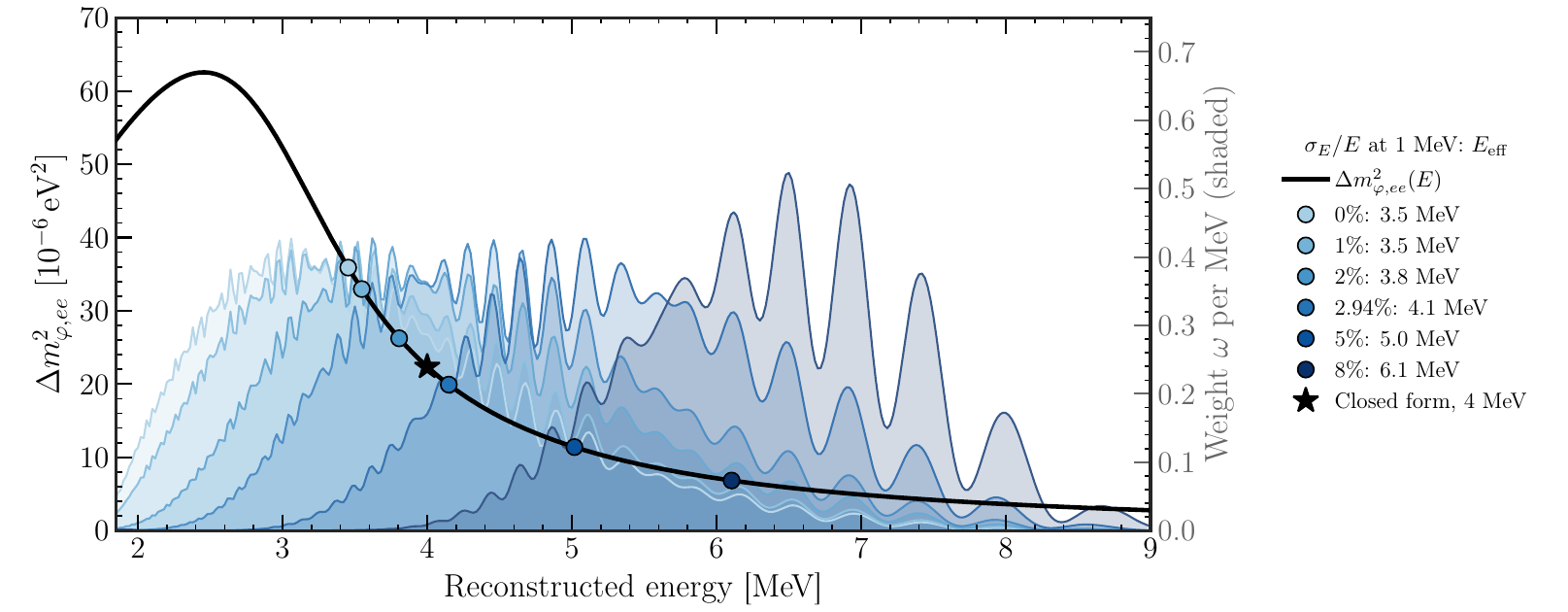}
\caption{JUNO, eight cores, in matter: the solar-phase term $\Delta m^2_{\varphi,ee}(E)$ (black line, left
axis), the weight $\omega_i$ of each reconstructed-energy bin for six values of the resolution $\sigma_E/E$ at 1~MeV from 0 to $8\%$ (shaded, right axis, smoothed, each with unit area, darker shades for worse resolutions), and the weighted average of $\Delta m^2_{\varphi,ee}$ for each resolution placed at its
effective energy (points, with $E_{\rm eff}$ in the legend). The star is the closed form at 4~MeV.}
\label{fig:juno_eeff}
\end{figure}

At the baseline of JUNO $\Delta_{21}$ is of order one, and the average of \meq{eq:mbr} depends on the weights. We define the effective energy $E_{\rm eff}$ as the energy at which $\Delta m^2_{\varphi,ee}(E)$ equals its weighted average $\avgw{\Delta m^2_{\varphi,ee}}$. Figure~\ref{fig:juno_eeff} shows $\Delta m^2_{\varphi,ee}(E)$, the weights of the bins for resolutions from 0 to 8\% at 1~MeV, and the average for each resolution placed at its effective energy. A worse resolution moves the weights to higher energies, and $E_{\rm eff}$ with them, from 3.5~MeV for a perfect resolution to 4.1~MeV at the predicted resolution of JUNO, 2.94\% at 1~MeV. The closed form with the solar phase at 4~MeV and $L=52.5$~km, $b=2s_{12}^2\dmsq{21}-\Delta m^2_{\varphi,ee}(4~{\rm MeV})$, therefore equals \meq{eq:mbr} only at the resolution for which $E_{\rm eff}=4$~MeV, about 2.6\% at 1~MeV. At the predicted resolution of JUNO the weighted average of \meq{eq:mbr} reproduces the full weighted bin average to within $1\times10^{-6}\eVsq$, and the closed form at 4~MeV lies $2\times10^{-6}\eVsq$ below the full weighted bin average. Bins of 10~keV instead of 20~keV change the intercept by $0.1\times10^{-6}\eVsq$, and bins of 100~keV, about twice the resolution near 3~MeV, raise it by $3\times10^{-6}\eVsq$. All values include matter effects.

\section{Atmospheric neutrinos in Earth matter}
\label{sec:s_atm}
This appendix supports \autoref{sec:atm}. It derives the matter shift above the MSW resonance and the suppression of the CP dependence. In the flavor basis the Hamiltonian is~\cite{Akhmedov:2004ny}
\begin{gather}
H=\frac{1}{2E}\Big[U\,{\rm diag}(0,\dmsq{21},\dmsq{31})\,U^\dagger+{\rm diag}(A,0,0)\Big],\nonumber\\
U=R_{23}\,\Gamma\,R_{13}\,\Gamma^\dagger R_{12},\qquad \Gamma={\rm diag}(1,1,e^{i\dCP}),
\end{gather}
where $R_{jk}$ is the rotation by $\theta_{jk}$ in the $j$--$k$ plane and
$A=2\sqrt2G_FN_eE$ for neutrinos ($-A$ for antineutrinos).

\subsection{Matter shift above the resonance}
\label{sec:s_shift}
\label{sec:s_decomp}
The difference $\delta_i=y_i-(-x+2c_{12}^2\dmsq{21})$ of \meq{eq:atm} contains four parts. The first is the matter shift of the atmospheric frequency. The second is the vacuum term
\begin{equation*}
(\kappa_\mu^{\NO}+\kappa_\mu^{\IO}-2c_{12}^2)\dmsq{21},
\end{equation*}
which is the CP term together with a small term of second order in $\sin\theta_{13}$. The third is the solar-phase term $-\Delta m^2_{\varphi,\mu\mu}$, and the fourth, when they are included, is the contribution of the other channels in the bin. The solar-phase term is negligible for DeepCore. Matter makes the $\delta_i$ differ between bins.

Above the resonance the $\nu_e$-like level lies far from the two levels of the $\nu_\mu$--$\nu_\tau$ system.
It shifts them only through its small mixing, at second order in perturbation theory
\cite{Akhmedov:2004ny,Minakata:2015gra,Denton:2016wmg}. In the $R_{23}$-rotated basis the
$\nu_e$-like level, near $A$, couples to the third level, at $c_{13}^2\dmsq{31}$, with strength
$c_{13}s_{13}\dmsq{31}$, and pushes it by $s_{13}^2(\dmsq{31})^2/(\dmsq{31}-A)$, to leading order in $s_{13}^2$. Its coupling to the second
level is proportional to $\dmsq{21}$ and negligible. For neutrinos in the normal ordering the third level is
pushed down, and the frequency $\Omega$ of the $\nu_\mu$--$\nu_\tau$ system, the difference of its two levels, decreases in size by $s_{13}^2x^2/(A-x)$. In the inverted ordering the third level, at about $-x$, is pushed away from the $\nu_e$-like level, and $|\Omega|$ increases by $s_{13}^2x^2/(A+x)$, to leading order with $|y|\simeq x$. Matching $|\Omega|$ between the orderings gives the difference $\delta_i$ of \meq{eq:shift_main}, with $+$ for neutrinos and $-$ for antineutrinos. For antineutrinos $A$ changes
sign, and $\hat A_i$ in \meq{eq:shift_main} denotes $|A|/x$ in both cases.

The first term of \meq{eq:shift_main}, $2s_{13}^2\cos2\theta_{12}\dmsq{21}/c_{13}^2\simeq1.3\times10^{-6}\eVsq$, is the difference between the solar term $2(c_{12}^2-s_{13}^2s_{12}^2)\dmsq{21}/c_{13}^2$ of the frequency above the resonance and $2c_{12}^2\dmsq{21}$, and does not depend on the matter density. The second term, the matter term, falls as $1/\hat A$ at high energy. At a fixed energy it grows as $x^2$ far above the resonance. The expansion
assumes that the $\nu_e$-like level is far from the third one, and it fails at the resonance, where the
second-order denominator vanishes. In the configuration of \autoref{fig:weights}, the bins between 20 and 50~GeV for up-going neutrinos, above the resonance, prefer $|y_i|$ between $2.421$ and $2.426\times10^{-3}\eVsq$, slightly above the solution, and the bins drawn in gray carry less than $0.05\%$ of the weight each.

To follow the shift through the resonance, where the expansion of \meq{eq:shift_main} fails, we use the exact eigenvalues of the two-level system formed by the $\nu_e$-like and the third level, with $\dmsq{21}$ set to zero. We evaluate them at the path-averaged PREM density of each cell of true energy and direction, and we average the shift with the weights of the detector. At the NuFIT phases this frequency shift reproduces about $70\%$ of $\datm$, the value quoted in \autoref{sec:atm_shift}. At $\dCP=90^\circ$, where the CP amplitude term of \eqr{eq:cp_amp} vanishes, it reproduces 72--77\% over the range of $x$. The rest is an amplitude effect. Near the resonance the effective $\sin^22\theta_{13}$ in matter approaches one, in neutrinos for one ordering and in antineutrinos for the other. The two orderings then differ also in the depth of the oscillation, and the IO fit compensates this difference with the frequency. The one-mass-scale probability in matter \cite{Akhmedov:2004ny}, with $\dmsq{21}$ set to zero, keeps both the frequency shift and the amplitude effect and reproduces 92--95\% of $\datm$ over the range of $x$ at $\dCP=90^\circ$. At the NuFIT phases a further $1.2$--$1.3\times10^{-6}\eVsq$ comes from the CP amplitude term of \eqr{eq:cp_amp}.

\subsection{Suppression of the CP-phase dependence in matter}
\label{sec:s_cp}
The suppression of the CP dependence in matter follows from the structure of the Hamiltonian given above. The matter potential commutes
with $R_{23}$ and with $\Gamma$, and $\Gamma$ commutes with $R_{12}$, so
$H=R_{23}\Gamma\,H_0\,\Gamma^\dagger R_{23}^{\rm T}$ with $H_0$ independent of
$\dCP$~\cite{Akhmedov:2008qt}. Two consequences follow, in vacuum and in matter alike
\cite{Kimura:2002wd,Denton:2016wmg}.
\begin{itemize}
\item The eigenvalues of $H$, and hence all oscillation frequencies, do not
depend on $\dCP$.
\item $\dCP$ enters $P_{\mu\mu}$ only through the projections of $\nu_\mu$ on the
eigenstates, $\langle\nu_\mu|v_i\rangle=c_{23}v_{i2}+s_{23}e^{i\dCP}v_{i3}$, with
$v_i$ the eigenvectors of $H_0$.
\end{itemize}

Above the MSW resonance, $|A|\gg|\dmsq{31}|$, the $\nu_e$-like eigenstate decouples \cite{Akhmedov:2004ny,Minakata:2015gra},
with a residual admixture of order $s_{13}\dmsq{31}/A$. The $\nu_\mu$ state then lives in a two-state
subspace, and $H_0$ restricted to the second and third states of the $R_{23}$-rotated basis is
\begin{equation}
2E\,H_{0}\big|_{23}=\begin{pmatrix}c_{12}^2\dmsq{21} & -s_{13}s_{12}c_{12}\dmsq{21}\\
-s_{13}s_{12}c_{12}\dmsq{21} & c_{13}^2\dmsq{31}+s_{13}^2s_{12}^2\dmsq{21}\end{pmatrix}
+\mathcal O\Big(\frac{s_{13}^2(\dmsq{31})^2}{A}\Big).
\end{equation}
This two-state system has a single frequency. The frequency is an eigenvalue difference and therefore does
not depend on $\dCP$, $\Omega\simeq c_{13}^2\dmsq{31}-(c_{12}^2-s_{13}^2s_{12}^2)\dmsq{21}$. Matching
$|\Omega|$ between the two orderings gives, for every bin,
\begin{equation}
y_i\simeq-x+\frac{2\,(c_{12}^2-s_{13}^2s_{12}^2)}{c_{13}^2}\,\dmsq{21}
=-x+2c_{12}^2\dmsq{21}+\mathcal O(s_{13}^2\dmsq{21}),
\end{equation}
the common part of \meq{eq:atm}, without any CP term.

The CP phase survives only in the amplitude of the oscillation. The small off-diagonal term of the
two-state Hamiltonian rotates its eigenstates by a small angle, of order $s_{13}\sin2\theta_{12}$ times
$\dmsq{21}/|\dmsq{31}|$. Through the phase $e^{i\dCP}$ in the projection of
$\nu_\mu$, this rotation changes the effective mixing angle of $\nu_\mu$ by an amount proportional to
$\cos\dCP$, and the probability becomes
\begin{equation}
P_{\mu\mu}\simeq1-\Big[1-\big(\cos2\theta_{23}\pm\zeta\big)^2\Big]\sin^2\frac{\Omega L}{4E},
\qquad \zeta=s_{13}\sin2\theta_{12}\sin2\theta_{23}\,\frac{\dmsq{21}}{|\dmsq{31}|}\,\cos\dCP,
\label{eq:cp_amp}
\end{equation}
with $+$ ($-$) for the normal (inverted) ordering, for neutrinos and antineutrinos alike. The
depth of the oscillation then changes with $\dCP$ by an amount proportional to $\zeta\cos2\theta_{23}$.
This change vanishes at maximal mixing, changes sign with the octant, and does not move the frequency. At finite $|A|$ the decoupled state contributes a third frequency with a weight of order
$s_{13}^2(\dmsq{31}/A)^2$, and through it a small dependence of the frequency on $\dCP$ returns. The same
admixture shifts $\Omega$ in opposite directions in the two orderings, which is the matter shift of
appendix~\ref{sec:s_shift}.

\section{Validation against the IceCube DeepCore data release}
\label{sec:s_valid}
This appendix describes the full fit of the DeepCore public data, the validation of the weighted bin average against it, the reproduction of the published
measurement, and the fit of both orderings to the observed data.

\subsection{Full-fit configuration}
\label{sec:s_fullfit}
Following the DeepCore oscillation analyses
\cite{IceCube:2017lak,IceCubeCollaboration:2023wtb,IceCubeCollaboration:2024ssx}, we build a standalone fitter for the
DeepCore public data release \cite{IceCube:B4RITM}, which provides the simulated events, the detector systematic
corrections, and the observed data. The full fit uses the expectation of the release for every flavor and interaction, both event classes of the release, the mixed and the track class, and the atmospheric-muon
background. The three-flavor oscillation probabilities are weighted by the flux of
the initial flavor of each event. The fit varies the thirteen parameters of \autoref{tab:fitpars}. The detector
parameters enter through the hypersurfaces of the release, which parameterize the
change of each bin with each detector parameter. The $\chi^2$ is the
modified Pearson form $\sum_i(N_i-n_i)^2/(N_i+\sigma^2_{{\rm MC},i})$, with $N_i$ the expected
and $n_i$ the observed counts of bin $i$, and $\sigma^2_{{\rm MC},i}$ the statistical variance of $N_i$ from the
finite number of simulated events. For the validation of appendix~\ref{sec:s_cmp} we build Asimov data at 25 values of $x$ in 2.2--2.8$\times10^{-3}\eVsq$, with the
NuFIT~6.0 oscillation parameters of \eqr{eq:inputs} and the nuisance parameters
at the best fit of the data of the release. The inverted ordering is fitted with all
thirteen parameters, with $\dCP=177^\circ$ in both orderings and the pull terms
centered at the true values. The hypersurfaces of the release depend on the splitting and are tabulated against $|\dmsq{31}|$ for NO oscillations. On Asimov data we evaluate them at the true splitting in both orderings, so that the detector correction is common to the data
and the fit, as in the weighted bin average.

\begin{table}[t]\centering
{\small\begin{tabular}{@{}ll@{}}
\toprule
Parameter & Prior\\
\midrule
$\sin^2\theta_{23}$, $\dmsq{32}$ & none\\
Optical efficiency & Gaussian, relative width $10\%$\\
Hole ice (two parameters) & none\\
Bulk-ice absorption and scattering & bounded scale factors, 0.90--1.10 and 0.95--1.15\\
Effective-area normalization & none\\
Atmospheric-muon normalization & none\\
Spectral index & Gaussian, absolute width $0.10$\\
Axial mass (two parameters) & Gaussian, width set by the parameterization of the release\\
Neutral-current normalization & Gaussian, relative width $20\%$\\
\bottomrule
\end{tabular}}
\caption{Parameters of the DeepCore full fit and their priors. The optical efficiency, the hole ice, and the bulk ice are the five detector parameters of the release.}
\label{tab:fitpars}
\end{table}

\subsection{Comparison with the weighted bin average}
\label{sec:s_cmp}
\begin{figure}[t]\centering
\includegraphics[width=0.66\textwidth]{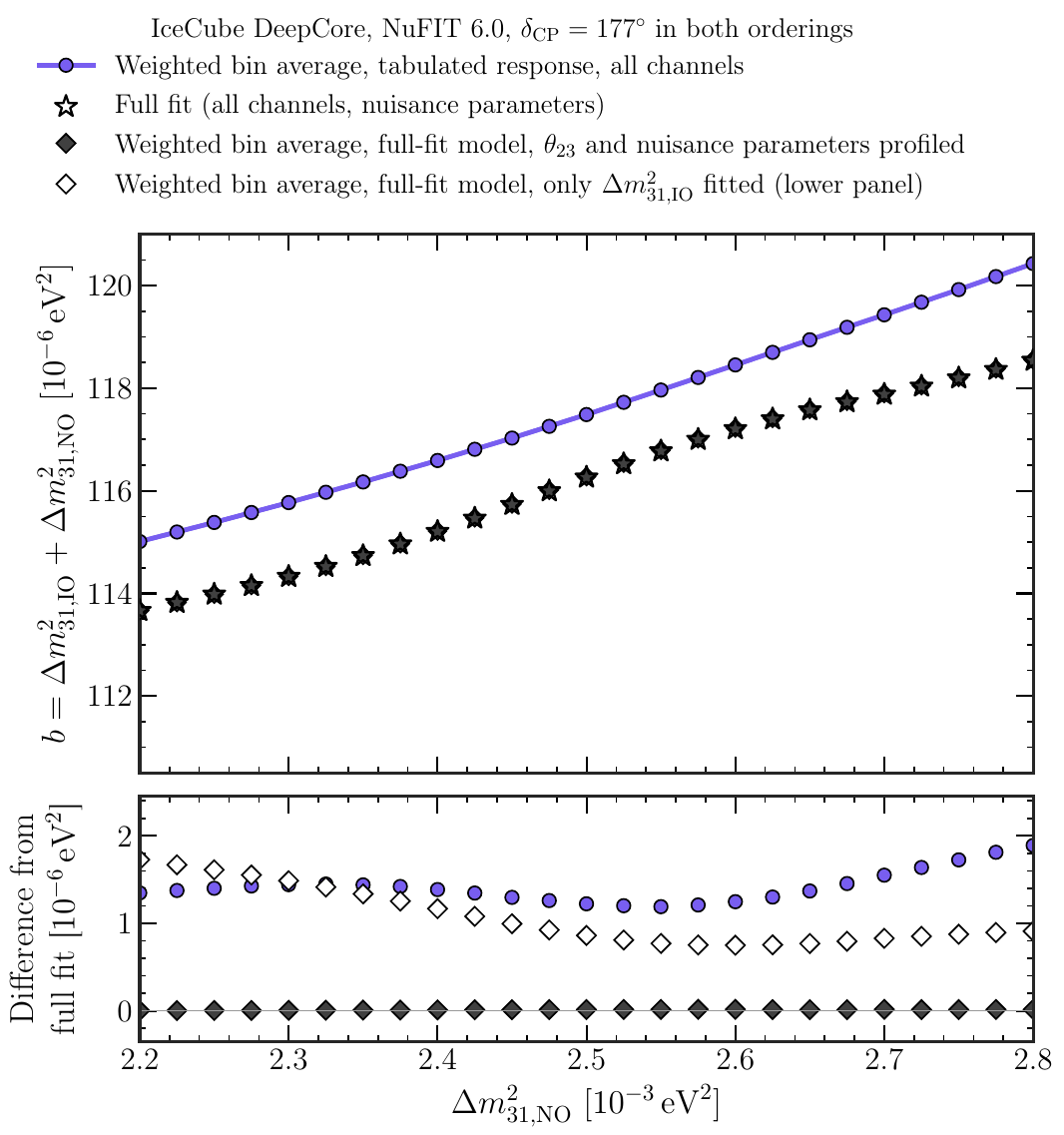}
\caption{Validation of the weighted bin average for IceCube DeepCore, NuFIT~6.0 with $\dCP=177^\circ$ in both
orderings. Top: intercept against the true $\dmsq{31,\NO}$ from the full fit (stars), from the weighted bin
average of \eqr{eq:master} on the tabulated response with all channels and $\theta_{23}$ profiled (line and circles), and on the expected
counts of the full fitter with $\theta_{23}$ and all nuisance parameters profiled (filled diamonds, which coincide with the stars).
Bottom:
difference of each evaluation from the full fit, including the weighted bin average
on the expected counts of the full fitter with only $\dmsq{31,\IO}$ fitted (open diamonds).}
\label{fig:valid}
\end{figure}

We fit the IO hypothesis to NO Asimov data at each true $\dmsq{31,\NO}$. As in the DeepCore analyses,
$\theta_{23}$ is profiled, while $\dCP$, to which DeepCore is hardly sensitive, is fixed. The comparison only requires the same phase in the fit and in the weighted bin average. This configuration differs from the reference configuration of \autoref{sec:atm}, so the intercepts differ from the value given there. Figure~\ref{fig:valid} compares the intercept of this full fit (stars) with
three evaluations of the weighted bin average. The first uses the tabulated response with all channels and
$\theta_{23}$ profiled (line and circles). The other two use the expected counts of the full fitter, which include
the detector model of the release, first with only $\dmsq{31,\IO}$ fitted (open diamonds) and then with
$\theta_{23}$ and all nuisance parameters profiled together with the splitting, by iterating
\eqr{eq:master_nuis} (filled diamonds). The lower panel shows the difference of each evaluation from the full
fit. Over the range of $x$ the three evaluations compare with the full fit as follows. The tabulated response with all channels and $\theta_{23}$ profiled lies above the full fit by $1.2$--$1.9\times10^{-6}\eVsq$, the difference quoted in \autoref{sec:atm_det}. The nuisance parameters account for $0.3$--$1.2\times10^{-6}\eVsq$ of this difference, and the approximations of the tabulated response for the rest. The weighted bin average on the expected counts of the full fitter, with only $\dmsq{31,\IO}$ fitted, lies above the full fit by at most $1.7\times10^{-6}\eVsq$. Since this evaluation differs from the full fit only in that $\theta_{23}$ and the nuisance parameters are not profiled, their effect is about a tenth of $\datm$. With $\theta_{23}$ and all nuisance parameters profiled, the iteration reaches the minimum of the full fit.

\subsection{Reproduction of the published DeepCore measurement}
\label{sec:s_repro}
\begin{figure}[t]\centering
\includegraphics[width=0.66\textwidth]{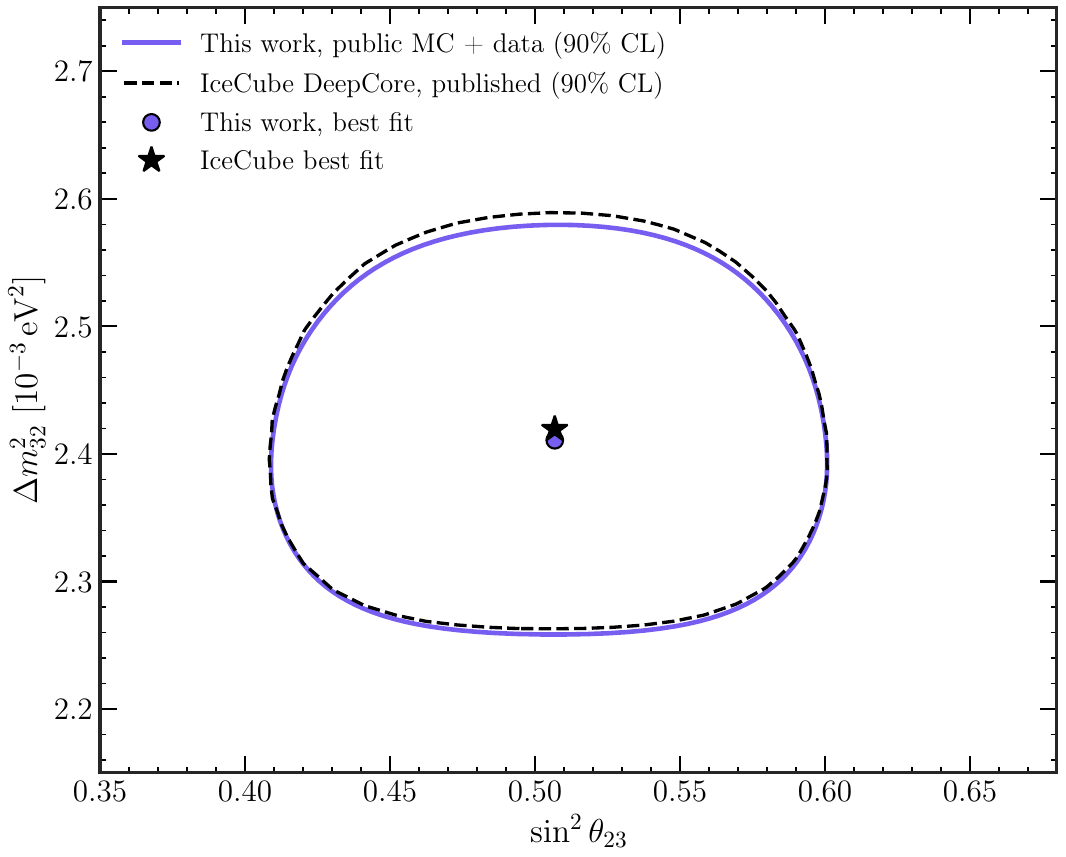}
\caption{The IceCube DeepCore measurement reproduced from the public release.
Solid: the 90\% contour of our fit to the public data, with every nuisance
parameter the release supports profiled with the priors of IceCube DeepCore. Dashed:
the published 90\% contour of IceCube DeepCore on the same sample, digitized from figure~26 of
ref.~\cite{IceCubeCollaboration:2023wtb}.
Dot: our best fit. Star: the published best fit, digitized from the same figure.}
\label{fig:icrepro}
\end{figure}

We fit the observed events of the release in normal ordering and compare the result with the IceCube DeepCore measurement on the same sample \cite{IceCubeCollaboration:2023wtb}. The fit uses the full expectation described above, $\sin^2\theta_{13}$ at the value of ref.~\cite{IceCubeCollaboration:2023wtb}, and the parameters of \autoref{tab:fitpars} with their priors. Figure~\ref{fig:icrepro} shows our
90\% contour and best fit together with the published ones, both contours assuming Wilks' theorem \cite{Wilks:1938dza}, while the published one-dimensional $68\%$ intervals are Feldman--Cousins. Our fit gives $\sin^2\theta_{23}=0.507\pm0.058$ and $\dmsq{32}=(2.411\pm0.078)\times10^{-3}\eVsq$, against the published $0.51\pm0.05$ and $(2.41\pm0.07)\times10^{-3}\eVsq$. The best fit and the
90\% contour agree with the published ones to a small fraction of a standard deviation, and our uncertainties,
from the curvature of the $\chi^2$ at its minimum, are within about $20\%$ of the published Feldman--Cousins
intervals.

\subsection{Fit of both orderings to the data}
\label{sec:s_iofit}
The detector hypersurfaces of appendix~\ref{sec:s_fullfit} are tabulated against the splitting of the NO simulation, so for the IO fit the splitting at which they are read is a choice. We read them at $|\dmsq{31,\IO}|+2c_{12}^2\dmsq{21}$. By \meq{eq:lbl} without the CP term, this is the NO splitting whose muon-disappearance pattern is closest to that of the IO model. The detector correction then follows the oscillation pattern rather than the numerical value of the splitting. On Asimov data at the NuFIT~6.0 best fit this reading reproduces the intercept obtained with the hypersurfaces at the true splitting to $2\times10^{-6}\eVsq$, while reading them at $|\dmsq{31,\IO}|$ lowers it by $14\times10^{-6}\eVsq$.

We apply the same fitter to the data of the release, with the fixed oscillation parameters of the release ($s_{12}^2=0.3098$, $\dmsq{21}=7.39\times10^{-5}\eVsq$, $\dCP=0$) and this reading. It returns $\dmsq{31}=\dmsq{32}+\dmsq{21}=(2.485\pm0.078)\times10^{-3}\eVsq$ under the normal and
$(-2.367\pm0.076)\times10^{-3}\eVsq$ under the inverted ordering. The intercept is $b=117.8\times10^{-6}\eVsq$. The two splittings move together
under statistical fluctuations, so $b$ is determined better than either splitting.

\section{Combination of experiments}
\label{sec:s_comb}
This appendix derives the synergy term of \meq{eq:dchi2}, its form for real data, where the two experiments measure different splittings, and its form with shared oscillation parameters, used in \autoref{sec:comb}.

\subsection{The synergy term}
For two independent experiments the combined IO $\chi^2$ is the sum of
the individual ones, minimized over the common parameters. The minimum of a sum is
not smaller than the sum of the minima. For Asimov data the NO minima of both
experiments and of their combination vanish, so
$\Delta\chi^2_{\rm NMO}\ge\Delta\chi^2_A+\Delta\chi^2_B$. The equality holds only
when both experiments return the same IO parameters. The excess is
the information carried by their disagreement, and it equals the parameter goodness-of-fit statistic of
refs.~\cite{Maltoni:2002xd,Maltoni:2003cu}, evaluated under the IO hypothesis.

For Asimov data the NO solutions of the two experiments coincide at the true splitting. Near its minimum the IO $\chi^2$ of
experiment $X$ is $\chi^2_{X,\min}+(y-f_X)^2/\sigma_X^2$. Here $\sigma_X$ is the width
of the IO fit in $y$, which is close to the NO measurement
error but not equal to it, and $f_A$ and $f_B$ are values of the same parameter,
$\dmsq{31,\IO}$. The sum is minimized over the common $y$ at
$y=(f_A/\sigma_A^2+f_B/\sigma_B^2)/(1/\sigma_A^2+1/\sigma_B^2)$, which leaves
$(f_A-f_B)^2/(\sigma_A^2+\sigma_B^2)$, the synergy term of \eqr{eq:dchi2}. At the common true splitting $f_A-f_B=b_A-b_B$ exactly, by the definition $b=f(x)+x$. In the joint fit each experiment is moved away from its own solution by $|\Delta b|\,\sigma_X^2/(\sigma_A^2+\sigma_B^2)$, and the parabolic form holds as long as this shift stays within the range where its IO $\chi^2$ is parabolic.

With real data the two experiments measure different values $x_A$ and $x_B$ under
the true ordering, with widths $\sigma_{A,\NO}$ and $\sigma_{B,\NO}$, and their
IO solutions are $f_X=-x_X+b_X$. Minimizing the combined $\chi^2$ in
each ordering in the same approximation gives
\begin{equation}
\Delta\chi^2_{\rm syn}=\frac{(\Delta b-\Delta x)^2}{\sigma_A^2+\sigma_B^2}-\frac{(\Delta x)^2}{\sigma_{A,\NO}^2+\sigma_{B,\NO}^2},
\qquad \Delta b=b_A-b_B,\quad \Delta x=x_A-x_B.
\label{eq:syn_data}
\end{equation}
This expression reduces to \eqr{eq:dchi2} for $\Delta x=0$. A difference $\Delta x$ of the measured splittings that moves the two IO best fits further apart enlarges the synergy term $\Delta\chi^2_{\rm syn}$, and one that moves them closer reduces it. The synergy term is negative when the two experiments agree better under the IO than under the NO. In this case the combination prefers the normal ordering less than the two experiments taken separately.

\subsection{Shared oscillation parameters}
When the two experiments also fit common parameters, such as $\theta_{23}$ or $\dCP^{\IO}$, the
IO solution of each experiment is a vector $\mathbf f_X$ with covariance $C_X$, twice the inverse of
the Hessian of its IO $\chi^2$ at the minimum, and the
same minimization gives
\begin{equation}
\Delta\chi^2_{\rm syn}=(\mathbf f_A-\mathbf f_B)^{\rm T}(C_A+C_B)^{-1}(\mathbf f_A-\mathbf f_B).
\label{eq:pair_general}
\end{equation}
The difference in $\dmsq{31}$ then counts together with the differences in the other parameters, all weighted by the inverse of $C_A+C_B$. In the flat valley of $\dCP^{\IO}$ the variance of an experiment in $\dCP^{\IO}$ is unbounded, the component of $\mathbf f_A-\mathbf f_B$ along the flat valley drops out, and the joint fit takes the value of $\dCP^{\IO}$ that brings the two splittings closest, as in \autoref{fig:synergy}.

In the left panel of \autoref{fig:synergy} the upper edge of the NOvA--DeepCore band comes from true values of $\sin^2\theta_{23}$ of $0.53$ or more, where the IO fit of NOvA returns the octant mirror of the true $\theta_{23}$ and that of DeepCore stays in the upper octant. In both panels, a joint fit of the two samples, which shares $\theta_{23}^{\IO}$ and $\dCP^{\IO}$, gives a larger synergy than \eqr{eq:dchi2} at almost all true values of $\dCP$ and $\theta_{23}$, and its largest values come from a true $\theta_{23}$ in the lower octant, where \eqr{eq:dchi2} gives nearly zero.

\providecommand{\href}[2]{#2}\begingroup\raggedright\endgroup

\end{document}